\documentclass[sigconf]{acmart}
\acmSubmissionID{379}

\usepackage{microtype}
\usepackage{graphicx}
\usepackage{subfigure}
\usepackage{booktabs} 
\usepackage{amsmath}
\usepackage{mathtools}
\usepackage{amsthm}
\usepackage{multirow}
\usepackage{xcolor}
\usepackage{colortbl}
\usepackage{bm}
\usepackage{array}
\usepackage{fancyhdr}
\usepackage{hyperref}
\usepackage{enumitem}
\usepackage{makecell} 
\usepackage[skip=2pt]{caption}
\usepackage[ruled,linesnumbered]{algorithm2e}
\AtBeginDocument{%
  }

\newcommand{\header}[1]{\vspace*{1mm}\noindent{\textbf{#1}}}
\author{Ziqi Zhao}
\orcid{0009-0008-3011-5745}
\affiliation{%
  \institution{Shandong University}
  \city{Qingdao}
  \country{China}
}
\email{ziqizhao.work@gmail.com}

\author{Zhaochun Ren}
\orcid{0000-0002-9076-6565}
\authornote{Co-corresponding authors.}
\affiliation{%
  \institution{Leiden University}
  \city{Leiden}
  \country{The Netherlands}
}
\email{z.ren@liacs.leidenuniv.nl}

\author{Jiyuan Yang}
\orcid{0000-0003-2700-5533}
\affiliation{%
  \institution{Shandong University}
  \city{Qingdao}
  \country{China}
}
\email{jiyuan.yang@mail.sdu.edu.cn}

\author{Zuming Yan}
\orcid{0009-0006-1387-901X}
\affiliation{%
  \institution{Shandong University}
  \city{Qingdao}
  \country{China}
}
\email{202200450040@mail.sdu.edu.cn}

\author{Zihan Wang}
\orcid{0000-0003-0493-2668}
\affiliation{%
  \institution{University of Amsterdam}
  \city{Amsterdam}
  \country{The Netherlands}
}
\email{zhw.cypher@gmail.com}

\author{Liu Yang}
\orcid{0009-0007-7508-0964}
\affiliation{%
  \institution{Shandong University}
  \city{Qingdao}
  \country{China}
}
\email{yangliushirry@gmail.com}

\author{Pengjie Ren}
\orcid{0000-0003-2964-6422}
\affiliation{%
  \institution{Shandong University}
  \city{Qingdao}
  \country{China}
}
\email{renpengjie@sdu.edu.cn}

\author{Zhumin Chen}
\orcid{0000-0003-4592-4074}
\affiliation{
    \institution{Shandong University}
    \city{Qingdao}
    \country{China}
}
\email{chenzhumin@sdu.edu.cn}

\author{Maarten de Rijke}
\orcid{0000-0002-1086-0202}
\affiliation{%
  \institution{University of Amsterdam}
  \city{Amsterdam}
  \country{The Netherlands}
}
\email{m.derijke@uva.nl}

\author{Xin Xin}
\orcid{0000-0001-6116-9115}
\authornotemark[1]
\affiliation{%
  \institution{Shandong University}
  \city{Qingdao}
  \country{China}
}
\email{xinxin@sdu.edu.cn}
\renewcommand{\shortauthors}{Ziqi Zhao et al.}
\setcopyright{rightsretained}
\copyrightyear{2025}
\acmYear{2025}
\acmDOI{}

\acmConference[SIGIR '25]{The 48th International ACM SIGIR Conference on Research and Development in Information Retrieval}{13 July -- 18 July, 2025}{Padua, Italy}

\acmISBN{}



\begin{document}

\title[Improving Sequential Recommenders through 
Counterfactual Augmentation of System Exposure]{Improving Sequential Recommenders through \\
Counterfactual Augmentation of System Exposure}

\begin{abstract}
In sequential recommendation (SR), system exposure refers to items that are exposed to the user. Typically, only a few of the exposed items would be interacted with by the user. Although SR has achieved great success in predicting future user interests, existing SR methods still fail to fully exploit system exposure data. Most methods only model items that have been interacted with, while the large volume of exposed but non-interacted items is overlooked. Even methods that consider the whole system exposure typically train the recommender using only the logged historical system exposure, without exploring unseen user interests.

In this paper, we propose \textbf{c}ounterfactual \textbf{a}ugmentation over \textbf{s}ystem \textbf{e}xposure for sequential \textbf{rec}ommendation (CaseRec).
To better model historical system exposure, CaseRec introduces reinforcement learning to account for different exposure rewards. CaseRec uses a decision transformer-based sequential model to take an exposure sequence as input and assigns different rewards according to the user feedback. 
To further explore unseen user interests, 
CaseRec proposes to perform counterfactual augmentation, where exposed original items are replaced with counterfactual items.
Then, a transformer-based user simulator is proposed to predict the user feedback reward for the augmented items. 
Augmentation, together with the user simulator, constructs counterfactual exposure sequences to uncover new user interests. 
Finally, CaseRec jointly uses the logged exposure sequences with the counterfactual exposure sequences to train a decision transformer-based sequential model for generating recommendation.  
Experiments on three real-world benchmarks show the effectiveness of CaseRec.
Our code is available at \url{https://github.com/ZiqiZhao1/CaseRec}.
\end{abstract}

\maketitle

\section{Introduction}

\begin{figure*}[t]
    \centering
        \subfigure[An illustrative example of system exposure.]{
        \begin{minipage}[t]{0.45\linewidth}
            \centering
            \includegraphics[width=0.9\columnwidth]{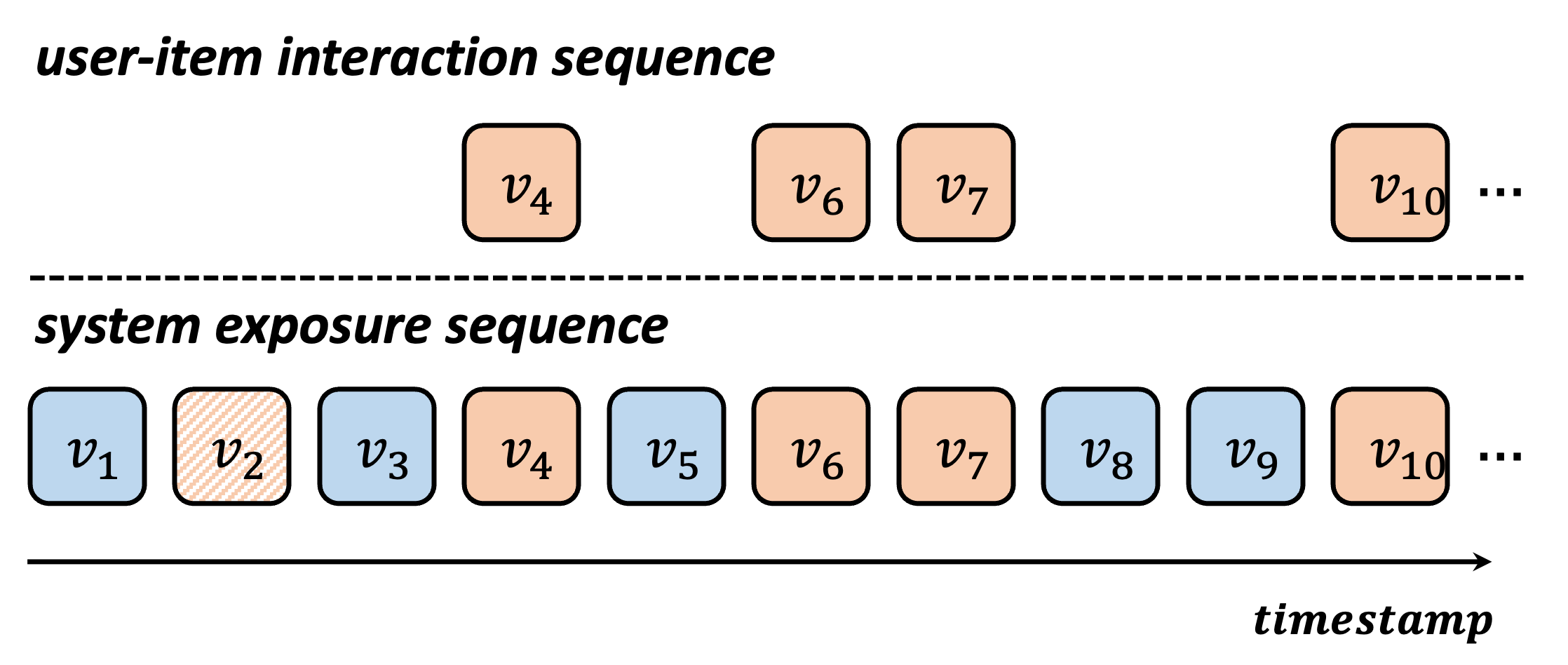}
        \end{minipage}
        \label{fig:system_exposure}
    }
    \subfigure[The statistical results of density across three datasets.]{
        \begin{minipage}[t]{0.45\linewidth}
            \centering
            \includegraphics[width=0.9\columnwidth]{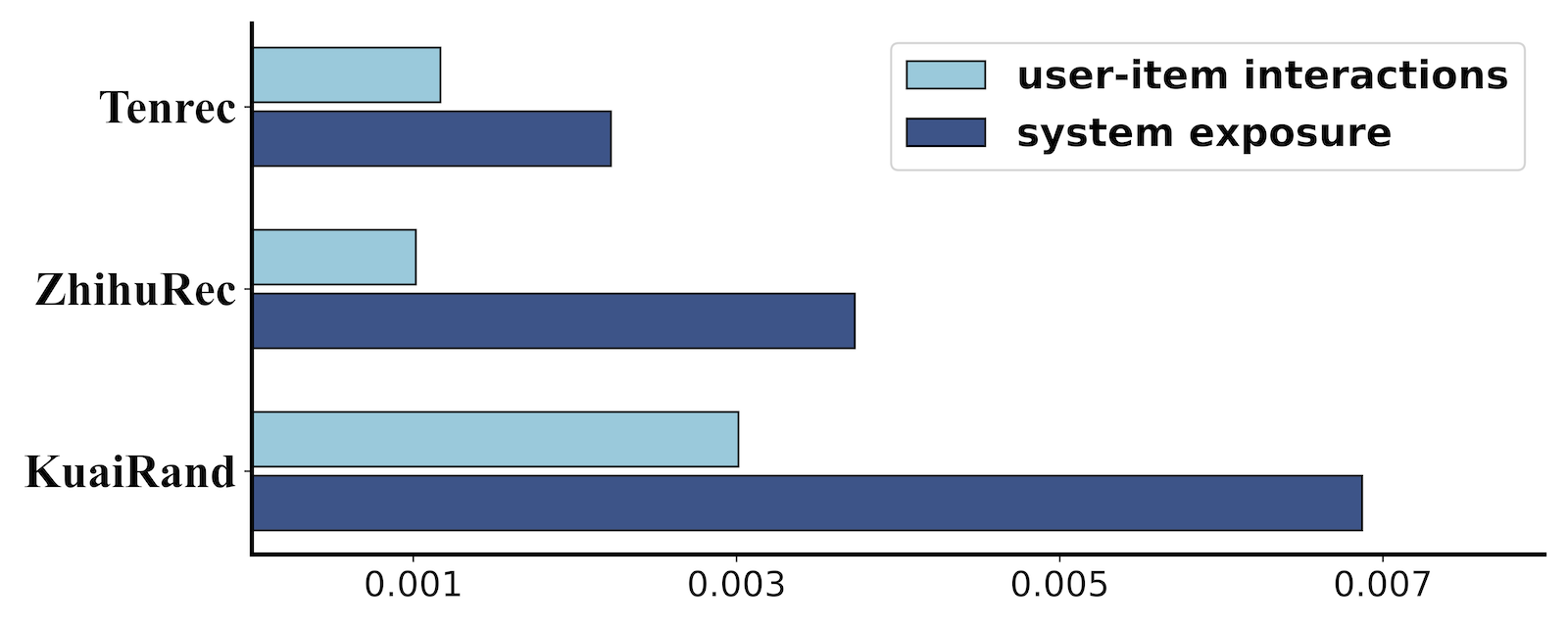}
        \end{minipage}
        \label{fig:exposure_density}
    }
    \caption{(a) Illustration of system exposure. Orange blocks refer to items interacted with by the user, 
    while blue blocks refer to items exposed by the system but not interacted with by the user. 
    The shadow $v_2$ denotes that $v_2$ could also interest the user but is not interacted with due to reasons like time limits.
    (b) System exposure is much larger than interacted data.}
\end{figure*}

Sequential recommendation (SR) aims to predict user future interests through modeling past user-item interaction sequences \cite{SASRec, zhou2020s3rec, reinforce-e-commerce, nextitnet}.
Unlike traditional collaborative filtering methods, SR models the user-item interactions as a dynamic sequence and uses (deep) sequential models to capture inherent sequential dependencies, enabling better recommendation \cite{sarwar2001itemcf, he2017ncf}.

\header{System exposure.} In the interaction process between users and the recommender system, the recommender exposes a personalized list of items, from which users select interesting items to interact with. 
Such a system behavior sequence is referred to as \emph{system exposure} \cite{DBLP:conf/www/expomf, yang2024debiasing}.
Figure \ref{fig:system_exposure} provides an illustrative example of system exposure. 
Figure \ref{fig:exposure_density} illustrates that users typically interact with only a few items from the exposed list. 
Generally speaking, system exposure is generated according to recommendation policies that represent the user preference from the point of view of the recommender system. 
Such exposure information can contribute to further improving future recommendation.

\header{Limitations of existing methods.} 
Although SR has achieved great success, existing SR methods still fail to fully exploit system exposure data, which limits their potential and leads to sub-optimal recommendation performance.

First, most SR methods only model items that have been interacted with, while the large volume of exposed but non-interacted items is overlooked. 
SR methods typically assume that exposed but non-interacted items denote negative user preference \cite{ding2018improved}. However, such a simple assumption may not always hold in practical scenarios. As an example in Figure~\ref{fig:system_exposure}, the shadow $v_2$ denotes that item $v_2$ could also interest the user but is not interacted with since the user can only select one item to interact with during a limited time period. 
This indicates that exposed but non-interacted items could also contain potential user interests. 
Besides, the large volume of exposure sequences also helps better learn sequential correlations between items, and thus provides more potential to further improve SR. From this perspective, we argue that exposure sequences, including both interacted and non-interacted items, should be modeled for SR.

Second, even if several studies that account for the whole system exposure, they only train the recommender using logged historical exposure and ignore those unseen user interests, leading to the increased exposure bias~\cite{biassurvey}.
Since training of existing SR models depends only on the logged exposure, previously deployed exposure policies may dramatically impact user-item interactions.
Exposure debiasing methods introduce penalties based on inverse propensity score (IPS) \cite{DBLP:conf/wsdm/SaitoYNSN20, liu2023bounding, 10.1145/3306618.3314288, Khalil2022cloze, wang2022lantentconfounders, DBLP:conf/cikm/Xu0CDW22, dai2022generalized,wang2019doubly} or distributionally robust optimization (DRO) \cite{yang2024debiasing}. 
However, introducing penalties on inappropriate items could negatively impact model performance. 
From this perspective, we argue that introducing counterfactual exposure sequences is an alternative solution to uncover new user interests and alleviate exposure bias.

\header{Proposed methods.} We propose \underline{\textbf{c}}ounterfactual \underline{\textbf{a}}ugmentation over \underline{\textbf{s}}ystem \underline{\textbf{e}}xposure for sequential \underline{\textbf{rec}}ommendation (CaseRec) to address the limitations mentioned above.

To better model the whole exposure sequence,  we need to consider different types of user feedback on exposed items. 
CaseRec introduces a reinforcement learning (RL)-based model to learn exposure sequences. 
RL has been proven to be effective in achieving reward-conditioned learning, which aligns well with modeling exposure sequences, i.e., by assigning different rewards according to different types of user feedback.
CaseRec uses a decision transformer (DT)-based sequential model to learn exposure sequences. 
DT casts the offline RL task as a sequence modeling problem and has the ability to handle complex long sequences. 
To adapt DT for modeling exposure sequences, CaseRec takes both the items and the corresponding user behavior as the input. Then, CaseRec introduces a high-dimensional encoder that combines item embeddings and corresponding behavior embeddings to ensure the differentiation between various user feedback over the exposure sequence. 
Finally, CaseRec redesigns the learning objective to ensure that the model can accurately predict the next item that the user is likely to interact with.

To augment counterfactual exposure sequences and uncover new user interests, CaseRec comes with two augmentation strategies, \textit{Random} and \textit{Self-Improving}, that replace part of the exposed items in a logged exposure sequence with other items.
Then, a transformer-based user simulator predicts the user feedback for the augmented items. 
The \textit{Random} strategy uses a simple yet effective uniform sampling strategy to replace items with randomly sampled new items, aiming at simulating user feedback under the random exposure policy and increasing the diversity of the training sequence. 
The \textit{Self-Improving} strategy conducts a small perturbation on one of historical exposed items, and uses the current SR model to generate the following exposure sequence. 
The generated exposure sequences are further used to train the SR model. 
This cycle allows the SR model to explore more of the item space near the historical exposed items.  
The counterfactual item replacement, together with the user simulator, constructs counterfactual exposure sequences to further uncover potential user interests. 
Finally, CaseRec uses logged exposure sequences and counterfactual augmented sequences to train the DT-based sequential model for generating recommendation.

To demonstrate the effectiveness of CaseRec, we conduct extensive experiments on benchmark datasets.
Experimental results show that CaseRec outperforms relevant state-of-the-art SR baselines and shows less exposure bias in the recommendation list.

\header{Main contributions.} To summarize, the main contributions of this work are the following:
\begin{itemize}[leftmargin=*,nosep]
\item We propose CaseRec, which uses an offline RL-based DT model to learn exposure sequences and account for various types of user feedback over exposed items.  
\item We propose two counterfactual augmentation strategies over system exposure to further uncover potential new user interests.
\item We propose a transformer-based user simulator to imitate user feedback for the augmented counterfactual items. 
\item Extensive experiments on three real-world datasets demonstrate the effectiveness of our methods for generating more accurate recommendation. CaseRec also shows promising performance in reducing exposure bias. 
\end{itemize}

\section{Preliminaries}

\begin{figure*}[ht]
\centering
\includegraphics[width=0.98\linewidth]{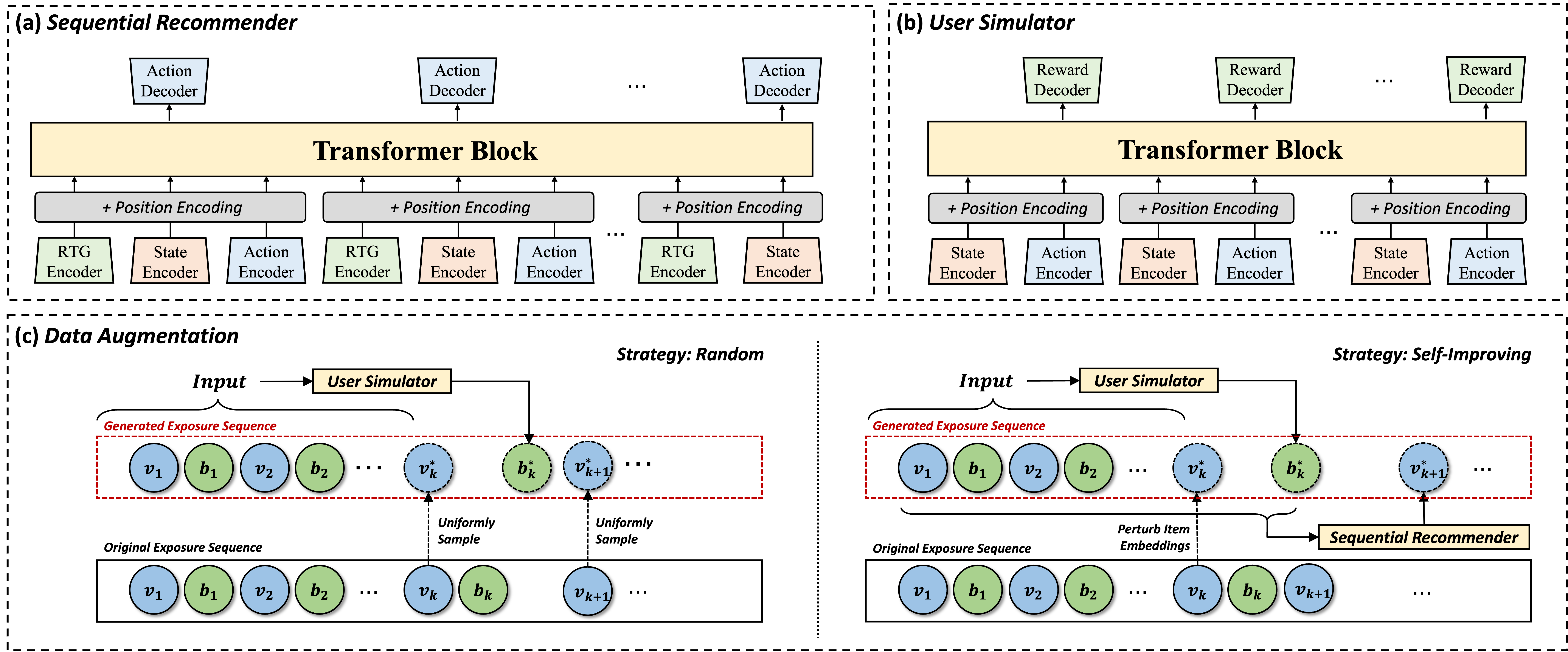}
\caption{
Overview of CaseRec. 
 (a) The architecture of DT-based sequential recommender, which takes system exposure as input and generates recommendation. Details can be found in Section \ref{sequtial_recommender}.
 (b) The architecture of transformer-based user simulator, which predicts user feedback for a given item. Details can be found in Section \ref{usersimulator}.
 (c) The counterfactual augmentation process, which illustrates two strategies: strategy \textit{Random} and strategy \textit{Self-Improving}. Details can be found in Section \ref{data_augmentation}.
The user simulator (b) is utilized to imitate user feedback for counterfactual items generated by the data augmentation (c), and the sequential recommender (a) is trained upon both logged exposure sequences and augmented counterfactual sequences. 
}
\label{fig:framework}
\end{figure*}

To provide a comprehensive understanding of the foundational concepts relevant to our study, we describe preliminaries about system exposure and offline reinforcement learning in this section.

\subsection{System exposure}
The items exposed to users by the recommendation agent are referred to as system exposure. 
Let $\mathcal{U}$ and $\mathcal{V}$ denote the user set and the item set, respectively, where $u \in \mathcal{U}$ denotes a user and $v \in \mathcal{V}$ denotes an item. 
Let $\mathcal{B}$ denote the user feedback to an exposed item (e.g., clicks, views or purchases). 
An $n$-length system exposure sequence of user $u$ can be denoted as  $S^u=(v^u_1,b^u_1,v^u_2,b^u_2,\ldots,v^u_n,b^u_n)$, where $v^u_j$ represents the $j$-th item exposed to user $u$, $b^u_j \in \mathcal{B}$ represents the feedback of user $u$ to item $v^u_j$. 
In this work, we consider $\mathcal{B}$ as a binary set, i.e., $\mathcal{B} = \{0,1\}$, where 1 denotes the item that has been interacted with by the user.  $\mathcal{B}$ can also contain more various user feedback for fine-grained user modeling. 

Most existing SR methods only consider interacted item sequences, i.e., items with $b^u_j=1$, while other items with $b^u_j=0$ are overlooked.  However, interacted items only account for a small portion of exposed items, and both interacted and non-interacted items should be modeled for better SR. To this end, we need to consider different user feedback over exposed items, which is accomplished through introducing RL-based models in this work.

\subsection{Offline reinforcement learning}
RL has achieved great success in conducting reward-driven decision-making. 
Different rewards can be assigned according to various user feedback, which naturally fits the modeling of exposure sequences.

A Markov decision process (MDP) is defined by the tuple $\mathcal{M}=(\mathcal{S},\mathcal{A},\mathit{P},r,\gamma)$, where $\mathcal{S},\mathcal{A}$ refer to state space and action space, respectively. 
Given a state $s_t \in \mathcal{S}$ and action $a_t \in \mathcal{A}$ at timestep $t$, $P(s_{t+1}|s_t,a_t)$ and $r(s_t,a_t)$ represent the distribution of next state $s_{t+1}$ and the obtained immediate reward, respectively. $\gamma \in (0,1)$ is the discount factor for future reward.
The learning objective of RL is to find an optimal policy that maximizes the expected cumulative rewards $ \mathbb{E}[\sum_{t}\gamma^{t}r(s_t,a_t)] $. 

Conventional RL algorithms collect the training data and learn the policy through online explorations with the environment (i.e., the user), which would affect the user experience when the policy is under-trained.
To solve this problem, we use offline RL methods in this work. Offline RL aims to learn the policy from an offline trajectory dataset pre-collected with unknown behavior policies, without online environment explorations.

Decision transformer is one of the most successful offline RL algorithms, which casts the offline RL task as a conditional sequence modeling problem. 
DT possesses both the capability to handle complex long sequences and the RL characteristic of reward-conditioned learning. 
Unlike traditional RL approaches that estimate value functions or compute policy gradients, 
DT auto-regressively generates desired future actions by modeling the trajectories of  
returns-to-go (RTG, i.e., expected future rewards), states $s$ and actions $a$. 
The input trajectory of DT is formulated as
\begin{equation}\label{eq1}
    \tau = (\hat{R}_{1},s_{1},a_{1},\ldots,\hat{R}_{T-1},s_{T-1},a_{T-1},\hat{R}_T,s_T),
\end{equation}
where $\hat{R}_t=\sum_{t}^{T}\gamma^tr_t$ is the RTG for timestamp $t$, representing the future cumulative reward from $t$ to $T$.
DT employs causally masked transformers, and the action $a_T$ is predicted through auto-regressive supervised learning over input trajectories.



\section{Methodology}

In this section, we present the details of CaseRec. 
Firstly we present the overview of the proposed framework.
Then, we describe each component of CaseRec individually.
The overall structure of CaseRec is provided in Figure \ref{fig:framework}.

\subsection{Model overview}\label{taskFormulation}
In this work, we consider adapting DT to model exposure sequences.
This is motivated by the following two reasons: (i) DT 
can handle complex long sequences to better learn item sequential correlations from exposure sequences; and (ii) DT can achieve reward-conditioned learning by assigning different rewards to different user feedback. 


Given the input trajectory $\tau_{rec} = (\hat{R}_{1},s_{1},a_{1},\ldots,\hat{R}_T,s_T)$, the elements in the trajectory are defined as:
\begin{itemize}[leftmargin=10pt]
    \item \textbf{state $s_t$ } represents the exposure sequence information before timestep $t$, which contains both the exposed the items and user's feedback towards the items at each timestep, i.e., 
\begin{equation}\label{eq2}
s_t=(v_1,b_1,v_2,b_2,\ldots,v_t,b_t)
\end{equation}
Besides, $s_t$ would be zero-padded or truncated to the length of $2L$, i.e., a total timestep of $L$ before $t$. To map the complex exposure sequence into a hidden state, we use a Gated Recurrent Unit (GRU)-based state encoder \cite{cho2014learning} as detailed in section \ref{Embeddingmodule}.
    \item \textbf{action $a_t\in\mathcal{V}$ } represents the item exposed to the user at timestep $t$, i.e., $a_t=v_{t+1}$. 
    \item \textbf{reward $r_t\in\mathbb{R}$ } represents the reward of the action $a_t$. We define the reward depending on the user feedback.
\begin{equation}\label{eq3}
r_t=
\begin{cases}
r_{uni} \text{, if } b_{t+1}=0\\
r_{int} \text{, if } b_{t+1}=1
\end{cases}
\end{equation}
where $r_{uni}$ and $r_{int}$ are hyperparameters which represent the corresponding rewards, respectively. Note that CaseRec can also support more flexible reward settings, e.g., assigning rewards according to the watching time for video recommendation.
    \item \textbf{RTG $\hat{R}_t \in \mathbb{R}$ } represents the cumulative reward from time $t$ through to time $T$.
\end{itemize}
The DT-based sequential recommender takes $\tau_{rec}$ as the input and predicts the action conditioned on $\hat{R}_t$ as the recommended items. 
Moreover, data augmentation is further introduced to augment counterfactual items for the exposure sequence, and the user simulator is used to predict user feedback towards counterfactual items. 
The data augmentation and the user simulator construct counterfactual exposure sequences to uncover potential new user interests. 
Finally, the sequential recommender is trained on the mixed dataset, which contains trajectories of both logged exposure sequences and counterfactual exposure sequences. 


\subsection{Sequential recommender}\label{sequtial_recommender}

In this section, we describe the details of the sequential recommender. 
\subsubsection{Encoding exposure sequences}\label{Embeddingmodule} For a given exposure sequence, the state encoder needs to not only capture the sequential signals but also model the signals of user behavior (i.e., feedback) towards the exposed items. 
To this end, we introduce two independent embedding layers to encode the item and the behavior, respectively. 
Subsequently, item embeddings and behavior embeddings are added together to combine the item information with the corresponding behavior information. 
Finally, the added embeddings are fed into a GRU model to capture the sequential signals and obtain the final state representation.

Specifically, given a state sequence $s_t=(v_1,b_1,v_2,b_2,\ldots,v_t,b_t)$, we first separate it into an item sequence $v_{1:t}=(v_1,v_2,\ldots,v_t)$ and a behavior sequence $b_{1:t}=(b_1,b_2,\ldots,b_t)$. 
The item embeddings and behavior embeddings can be defined as:
\begin{equation}\label{eq4}
\mathbf{v}_i = \text{E}_{item}(v_i)\text{ , } \forall v_i \in v_{1:t}
\end{equation}
\begin{equation}
\mathbf{b}_i = \text{E}_{behavior}(b_i)\text{ , } \forall b_i \in b_{1:t}
\end{equation}
where $\text{E}_{item}$ and $\text{E}_{behavior}$ represent the embedding functions, which could be achieved by simply looking up trainable embedding tables. $\mathbf{v}_i \in \mathbb{R}^d$ and $\mathbf{b}_i \in \mathbb{R}^d$ represent the item embedding and the behavior embedding respectively, where $d$ denotes the embedding size. 
Then the added embedding can be obtained by
\begin{equation}
\mathbf{x}_i = \mathbf{v}_i+\mathbf{b}_i\text{ , } \forall i \in [1,t].
\end{equation}
Consequently, the state embedding $\mathbf{s}_t$ for a given sequence $s_t$ is obtained from a GRU model as
%
\begin{equation}
\label{eq:gru}
\mathbf{s}_t=\text{GRU}(\mathbf{x}_t,\mathbf{s}_{t-1})
\end{equation}
Note that the GRU model can also be replaced with other sequential models. For simplicity, in this paper, we use the GRU model as a default encoder.

Besides, we also introduce an RTG encoder and an action encoder to embed RTG and actions. The representations for RTG $\hat{R}_t$ and action $a_t$ are defined as:
\begin{align}
    \mathbf{e}_{\hat{R}_t} & = \tanh(\text{E}_{RTG}(\hat{R}_t))
    \\
    \mathbf{a}_t & = \tanh(\text{E}_{item}(a_t)),
    \label{eq9}
\end{align}
where $\text{E}_{RTG}$ and $\text{E}_{item}$ represent the embedding functions of $\hat{R}_t$ and $a_t$, respectively. In our implementation, the same embedding table of item is used for both Eq.\ref{eq9} of the action encoder and Eq.~\ref{eq4} of the state encoder, allowing the embedding table to be fully trained. 
Then the representation of each tuple at timestep $t$ can be formulated as $[\mathbf{e}_{\hat{R}_t},\mathbf{s}_t,\mathbf{a}_t] \in \mathbb{R}^{3 \times d}$. Additionally, a position embedding of timestep $t$ is learned and added to each embedding, yielding the representation 
$[{\mathbf{e}}_{\hat{R}_t}',{\mathbf{s}}_t',{\mathbf{a}}_t']$.
The final representation of the input trajectory is defined as
\begin{equation}
\boldsymbol{\tau}_{rec} = ({\mathbf{e}}_{\hat{R}_1}',{\mathbf{s}}_1',{\mathbf{a}}_1',\ldots,{\mathbf{e}}_{\hat{R}_{T-1}}',{\mathbf{s}}_{T-1}',{\mathbf{a}}_{T-1}',{\mathbf{e}}_{\hat{R}_{T}}',{\mathbf{s}}_{T}').
\end{equation}

\subsubsection{Supervised learning} The primary model architecture is based on a unidirectional transformer layer, incorporating a multi-head self-attention mechanism~\cite{vaswani2017attention}. The output representation of trajectory is defined as
\begin{equation}
\hat{\boldsymbol{\tau}}_{rec} = \text{FFN}[\text{MHA}(\boldsymbol{\tau}_{rec})]=(\hat{\mathbf{e}}_{\hat{R}_1},\hat{\mathbf{s}}_1,\hat{\mathbf{a}}_1,\ldots,\hat{\mathbf{a}}_{T-1},\hat{\mathbf{e}}_{\hat{R}_{T}},\hat{\mathbf{s}}_{T}).
\end{equation}
$\text{MHA}$ is a multi-head self-attentive layer and $\text{FFN}$ is feed-forward neural layers with with GELU \cite{hendrycks2016gaussian} activation function and skip-connection. To avoid overfitting and enable a more stable learning without vanishing or exploding gradient issues, we include dropout layers and layer normalization \cite{ba2016layer}. 
The representation $\hat{\mathbf{s}}_t$ is fed into a fully connected layer to compute the classification logits on candidate actions for the $t$-th timestep as 
\begin{equation}
\tilde{\mathbf{a}}_t=[y_1^t,y_2^t,\ldots,y_{|\mathcal{V}|}^t]=\mathbf{W}_i\hat{\mathbf{s}}_t+\mathbf{c}\text{ , } \forall t \in [1,T],
\end{equation}
where $\tilde{\mathbf{a}}_t$ is predicted scores for candidate actions at timestep $t$. $\mathbf{W}_i \in \mathbb{R}^{n \times d}$ and $\mathbf{c} \in \mathbb{R}^{n}$ are trainable parameters.

The learning objective of the original DT is to predict the action $a_t$ at timestep $t$, 
while the learning objective of SR is to predict the next item that the user is likely to interact with based on the previously exposed items and user feedback. 
In exposure sequences, there are both interacted and non-interacted items, so directly using the auto-regressive objective of DT would lead to the gap between training objectives.
To this end, we reformulate the learning objective of CaseRec to predict only high-rewarded actions, i.e., the item that the user is likely to interact with, based on the exposed items prior to timestep $t$. 
Therefore, the ground-truth action for $\tilde{\mathbf{a}}_t$ is defined as the next item that the user would interact with:
\begin{equation}
    \bar{a}_t = v_k \text{ , where } k=\text{min}\{k \mid k>t \wedge b_k = 1\}
\end{equation}
Then the cross-entropy loss between predicted actions and ground-truth actions is used to train the model:
\begin{equation}\label{eq14}
\mathcal{L}_{rec}= \frac{1}{T}\sum_{t=1}^T \sum_{i=1}^{|\mathcal{V}|} Y^t_i \log(p_i^t), \text{ where } p_i^t=\frac{e^{y_i^t}}{\sum_{j\in\mathcal{V}}e^{y_{j}^t}}.
\end{equation}
$Y^t_i$ is the binary indicator and equals to 1 if the $i$-th item is the ground-truth $\bar{a}_t$, otherwise $Y^t_i=0$. 
In the training stage, the RTG can be obtained by calculating the sum of future rewards. 
In the inference stage, we can obtain the RTG through auto-regressive generation. 
For example, in practical online deployments, we can set the expected RTG at the beginning and then decrease the RTG at each timestep according to the real user feedback. 
For offline inference and evaluation, we typically set the
discounter factor as $\gamma=1$, and the RTG at timestep $t$ is set as $\hat{R}_t=\sum_{k=t}^{T-1} r_k + r_{int}$, which means that the sequential recommender is expected to generate an action that can be interacted by the user.


\subsection{Counterfactual augmentation}

To further explore unseen user interests, we proposed two counterfactual augmentation strategies, namely strategy \textit{Random} and strategy \textit{Self-Improving}, to generate additional exposure sequences.
The main difference between the two strategies lies in the method of replacing items in the original exposure sequence.
Specifically, strategy \textit{Random} replaced the original items with uniformly sampled items to alleviate the effect of previously deployed recommendation policies.
On the other hand,  strategy \textit{Self-Improving} firstly replaced the original item by perturbing the item embedding, and then 
generated the following exposure sequence through the current sequential recommender in an auto-regressive manner to uncover new user interests.
We intend to investigate \emph{what's the user feedback if a counterfactual item is exposed to the user?}
To this end, we introduce a user simulator to predict the user feedback for the given counterfactual items and conduct counterfactual augmentation by replacing items in the logged exposure sequences and predicting corresponding user feedback.

\subsubsection{User simulator}\label{usersimulator}
Motivated by recent offline model-based reinforcement learning, which improves policy learning by using environment models to further explore the state-action space\cite{yu2021combo,kidambi2020morel,yu2020mopo},  the user simulator is proposed to predict the user feedback reward for counterfactual items.
Considering that the exposure sequence is complex and informative, the transformer architecture, with outstanding long sequence modeling capability and generalization capability, is utilized to predict user feedback reward. 

Specifically, given an input sequence of user simulator $(s_{1}$, $a_{1}$, $s_{2}$, $a_{2}$, \ldots, $s_{T}$, $a_{T})$, the goal of user simulator is to predict the user feedback reward $r_{t}$ of action $a_{t}$. 
The user simulator and the sequential recommender employ similar components, including the state encoder, the action encoder, and transformer blocks. 
The output representation at position $a_t$ from the transformer block is fed into a fully connected layer with the binary cross-entropy loss function to optimize the simulator parameters. 
At the inference stage, the generated reward at last timestep $T$ is used to predict the user feedback reward $r_T$ of action $a_T$.

\begin{algorithm}[t]
	\caption{Training procedure of CaseRec.}
	\label{alg:algorithm1}
	\KwIn{sequential recommender $\mathcal{M}$, user simulator $\mathcal{G}$, augmentation ratio $\delta$, generation length $h$, number of max epochs $k$, original dataset $\mathcal{D}$, item set $\mathcal{V}$}
	\BlankLine
	Train $\mathcal{M}$ with original dataset $\mathcal{D}$
    
        \For{$i$=\textnormal{1 to }$k$}{
            Initialized augmentation set $\mathcal{D}_{aug} = \varnothing$
            
            \While{$|\mathcal{D}_{aug}|/|\mathcal{D}|<\delta$}{
                Sample $s_{ori} = \{v_i,b_i\}_{i=1}^{k-1}$ and $v_k$ from $\mathcal{D}$

                \uIf{\textit{Self-Improving}}{
                
                    Get $v_k^*$ by perturbing the embeddings of $v_k$\
                }

                \uElseIf{\textit{Random}}{
                
                    Get $v_k^*$ by uniformly sampling from $\mathcal{V}$\
                }

                Obtain $b_k^*$ from $\mathcal{G}$ based on $s_{ori} \cup v_k^*$

                $s_{ori}=s_{ori} \cup \{ v_k^*, b_k^*\}$

                \For{$t$=$k+1$\textnormal{ to }$k+h-1$}{

                    \uIf{\textit{Self-Improving}}{
                
                        Obtain $v_t^*$ from $\mathcal{M}$ based on $s_{ori}$}\
                    \uElseIf{\textit{Random}}{
                
                        Get $v_t^*$ by uniformly sampling from $\mathcal{V}$}\
                    Obtain $b_t^*$ from $\mathcal{G}$ based on $s_{ori} \cup v_t^*$
                    
                    $s_{ori}=s_{ori} \cup \{ v_t^*, b_t^*\}$
                }

                $\mathcal{D}_{aug}$.append($s_{ori}$)
            }

            Train $\mathcal{M}$ with augment dataset $\mathcal{D}_{aug}$
        }
\end{algorithm}

\subsubsection{Data augmentation}\label{data_augmentation}
In this section, we present the two data augmentation strategies individually.

\textbf{Strategy \textit{Random}.} 
Recent works have shown that current recommendation models may be influenced by prior recommendation policies, resulting in sub-optimal recommendation performance\cite{yang2024debiasing,biassurvey}.
To prevent the augmented sequences from being influenced by previously deployed recommendation policies, we perform random exposure simulation to generate counterfactual exposure sequences.
Specifically, given a logged exposure sequence from the original training data $s_{ori} = (v_1,b_1,v_2,b_2,\ldots,v_T,b_T)$, we replace the item $v_k$ with a uniformly sampled new item $v_k^*$, where $k \in [1,T]$. 
Then the sequence $(s_{1},a_{1},s_{2},a_{2},\ldots ,s_{k-1},v_k^*)$ can be constructed from $(v_1,b_1,v_2,b_2,\ldots,v_k^*)$ and fed into the user simulator to predict the user feedback $r_{k-1}^*$, i.e., $b_k^*$ towards the new item $v_k^*$.
This process is repeated for $h$ times to generate a counterfactual exposure sequence as $s_{gen} = (v_1,b_1,\ldots,v_k^*,b_k^*,\ldots,v_{k+h-1}^*,b_{k+h-1}^*)$.
After data augmentation, the sequential recommender is trained on trajectories generated from both logged exposure sequences and counterfactual exposure sequences through Eq.\ref{eq14}.

\textbf{Strategy \textit{Self-Improving}.}
Inspired by the RL agent's improvement of its policy through continuous interaction with the environment, the \textit{Self-Improving} strategy leverages the current sequential recommender to generate out-of-distribution data in an auto-regressive manner, subsequently utilizing the data to enhance the sequential recommender further.
Specifically, given a logged sequence $s_{ori} = (v_1,b_1,v_2,b_2,\ldots,v_k)$ from the original training dataset, we first add a small Gaussian noise to the item embedding of $v_k$ to obtain new item embedding, and then replace item $v_k$ with the $v_k^*$, whose embedding exhibits the highest cosine similarity to the newly perturbed embedding.
The corresponding sequence is then fed into the user simulator to predict the user feedback $b_k^*$ towards the new item.
Subsequently, based on the exposure sequence, we use the sequential recommender, pre-trained on the original dataset, to generate the next item and the user simulator to predict the corresponding user feedback.
Similar to strategy \textit{Random}, this process is also repeated to generate a new sequence with a length of $h$.
After augmenting a specific number of data, these data will be used to update the parameters of the sequential recommender.
The entire loop will continue until the number of max epochs is reached or the early stopping is triggered.
Algorithm \ref{alg:algorithm1} details the training procedure of CaseRec with two augmentation strategies.


\section{Experiments}
In this section, we perform experiments on three real-world datasets to verify the effectiveness of CaseRec.
We aim to answer the following research questions:
\begin{itemize}[leftmargin=10pt]
\item \textbf{RQ1:} How does the recommendation performance of CaseRec compare with other sequential recommendation baselines?
\item \textbf{RQ2:} How does the debiasing performance of CaseRec compare with other debiasing methods?
\item \textbf{RQ3:} How does the counterfactual augmentation influence the performance of CaseRec?
\item \textbf{RQ4:} How does the design of CaseRec affect the recommendation performance?
\end{itemize}
\subsection{Experimental settings}

\begin{table*}[t]
    \caption{Overall performance comparison of different methods on the three datasets. NG and Re is short for NDCG and Recall, respectively. '-R' and '-S' represent the \textit{Random} strategy and \textit{Self-Improving} strategy, respectively. Boldface denotes the highest score. * denotes the significance $p$-value $<0.01$ compared with the best baseline which is marked with \underline{underline}.}
    \centering
    \setlength{\tabcolsep}{2pt}
    \resizebox{\linewidth}{!}
    {%
        \begin{tabular}{ll ccccc ccccccc}
        \toprule
         &  & \multicolumn{11}{c}{\textbf{Method}} \\
        \cmidrule{3-14}
         \textbf{Dataset} & \textbf{Metric} & GRU4Rec & { SASRec } & BERT4Rec & { CORE } & { FEARec } & { IPS}  & { RelMF} & { DRO} & { DT } & { SQN} & \textbf{CaseRec-R}& \textbf{CaseRec-S} \\
        \midrule
        \multirow{6}{*}{\textbf{ZhihuRec}} 
                                           & Re@5  &  0.0113  &  0.0063  &  \underline{0.0138}  &  0.0088  &  0.0137  &  0.0090  &  0.0101  &  0.0101  &  0.0126  & 0.0095 &  {0.0528}$^*$ & \textbf{0.0704}$^*$  \\
                                           & Re@10 &  0.0214  &  0.0201  &  0.0176  &  0.0163  &  \underline{0.0244}  &  0.0189  &  0.0138  &  0.0151  &  0.0201  & 0.0232 &  {0.0741}$^*$ & \textbf{0.0942}$^*$  \\
                                           & Re@20 &  0.0251  &  0.0402  &  0.0289  &  0.0402  &  \underline{0.0445}  &  0.0255  &  0.0239  &  0.0264  &  0.0437  & 0.0432 &  {0.1131}$^*$ & \textbf{0.1470}$^*$  \\
                                           & NG@5  &  \underline{0.0090}  &  0.0032  &  0.0085  &  0.0050  &  0.0083  &  0.0056  &  0.0075  &  0.0076  &  0.0077  & 0.0054 &  {0.0352}$^*$ & \textbf{0.0464}$^*$  \\
                                           & NG@10 &  \underline{0.0123}  &  0.0076  &  0.0097  &  0.0074  &  0.0117  &  0.0086  &  0.0087  &  0.0092  &  0.0101  & 0.0087 &  {0.0421}$^*$ & \textbf{0.0540}$^*$  \\
                                           & NG@20 &  0.0132  &  0.0125  &  0.0126  &  0.0134  &  \underline{0.0167}  &  0.0102  &  0.0112  &  0.0120  &  0.0159  & 0.0143 &  {0.0519}$^*$ & \textbf{0.0672}$^*$  \\
        \midrule
        \multirow{6}{*}{\textbf{Tenrec}}   
                                           & Re@5  &  0.0574  &  0.0571  &  0.0445  &  0.0463  &  0.0432  &  0.0460  &  0.0508  &  \underline{0.0627}  &  0.0600  & 0.0581 &  {0.0895}$^*$ & \textbf{0.1100}$^*$  \\
                                           & Re@10 &  0.0987  &  0.0946  &  0.0848  &  0.0867  &  0.0771  &  0.0706  &  0.0687  &  \underline{0.1100} &  0.1044  & 0.1054 &  {0.1419}$^*$ & \textbf{0.1624}$^*$  \\
                                           & Re@20 &  0.1608  &  0.1526  &  0.1463  &  0.1491  &  0.1112  &  0.0971  &  0.0980  &  \underline{0.1709}  &  0.1701  & 0.1687 &  {0.2172}$^*$ & \textbf{0.2459}$^*$  \\
                                           & NG@5  &  0.0362  &  0.0362  &  0.0299  &  0.0276  &  0.0258  &  0.0309  &  0.0325  &  \underline{0.0391}  &  0.0353  & 0.0378 &  {0.0566}$^*$ & \textbf{0.0719}$^*$  \\
                                           & NG@10 &  0.0494  &  0.0484  &  0.0428  &  0.0405  &  0.0385  &  0.0388  &  0.0382  &  \underline{0.0542}  &  0.0495  & 0.0512 &  {0.0735}$^*$ & \textbf{0.0888}$^*$  \\
                                           & NG@20 &  0.0650  &  0.0630  &  0.0583  &  0.0562  &  0.0471  &  0.0455  &  0.0456  &  \underline{0.0695}  &  0.0661  & 0.0676 &  {0.0925}$^*$ & \textbf{0.1099}$^*$  \\
        \midrule
        \multirow{6}{*}{\makecell{\textbf{KuaiRand} \\ \textbf{-15 policies}}} 
                                           & Re@5  &  0.0675  &  0.0600  &  0.0518  &  0.0576  &  0.0358  &  0.0571  &  0.0327  &  0.0584  &  0.0644  & \underline{0.0681} &  {0.1333}$^*$ & \textbf{0.1644}$^*$  \\
                                           & Re@10 &  \underline{0.1101}  &  0.1039  &  0.0874  &  0.0948  &  0.0575  &  0.0973  &  0.0580  &  0.1093  &  0.1042  & 0.1079 &  {0.2000}$^*$ & \textbf{0.2311}$^*$  \\
                                           & Re@20 &  0.1706  &  0.1652  &  0.1474  &  0.1710  &  0.0962  &  0.1586  &  0.0936  &  \underline{0.1718}  &  0.1634  & 0.1574 &  {0.2948}$^*$ & \textbf{0.3217}$^*$  \\
                                           & NG@5  &  \underline{0.0428} &  0.0359  &  0.0309  &  0.0333  &  0.0231  &  0.0335  &  0.0206  &  0.0350   &  0.0406  & 0.0419 &  {0.0876}$^*$ & \textbf{0.1135}$^*$  \\
                                           & NG@10 &  \underline{0.0566}  &  0.0499  &  0.0423  &  0.0453  &  0.0301  &  0.0465  &  0.0288  &  0.0513  &  0.0507  & 0.0545 &  {0.1088}$^*$ & \textbf{0.1350}$^*$  \\
                                           & NG@20 &  0.0718  &  0.0651  &  0.0574  &  0.0645  &  0.0397  &  0.0619  &  0.0377  &  \underline{0.0770}   &  0.0706  & 0.0720 &  {0.1326}$^*$ & \textbf{0.1579}$^*$  \\
        
        \bottomrule
        \end{tabular}
    }
    \label{tab:main_results}
    \vspace{-10pt}
\end{table*}
    
        

\begin{table}[ht]
    \caption{Exposure debiasing performance on the KuaiRand-Random dataset. NG and Re is short for NDCG and Recall, respectively. `-R' and `-S' represent the \textit{Random} strategy and \textit{Self-Improving} strategy, respectively. Boldface denotes the highest score. * denotes the significance $p$-value $<0.01$ compared with the best baseline which is marked with \underline{underline}.}
    \centering
    \setlength{\tabcolsep}{1mm}
    \resizebox{\linewidth}{!}
    {%
        \begin{tabular}{l ccc ccc}
        \toprule
        & \multicolumn{6}{c}{\textbf{Metric}} \\
        \cmidrule{2-7}
        \textbf{Method} & Re@5 & Rec10 & Re@20 & NG@5 & NG@10 & NG@20\\
        \midrule        
        GRU4Rec &0.0012&0.0020&0.0050&0.0007&0.0010&\underline{0.0018} \\
        SASRec &0.0011&0.0026&0.0052&0.0006&0.0011&0.0017 \\
        SQN &0.0012&\underline{0.0028}&0.0050&0.0007&\underline{0.0013}&0.0016\\
        IPS &0.0012&0.0022&0.0040&0.0007&0.0010&0.0015 \\
        RelMF &0.0010&0.0023&\underline{0.0055}&0.0005&0.0010&\underline{0.0018}\\
        DRO & \underline{0.0014} & 0.0022 & 0.0046 & \underline{0.0008} & 0.0010 & 0.0016 \\
        \textbf{CaseRec-R} & \textbf{0.0085}\rlap{$^*$} & \textbf{0.0139}\rlap{$^*$} & \textbf{0.0229}\rlap{$^*$} & \textbf{0.0059}\rlap{$^*$} & \textbf{0.0077}\rlap{$^*$} & \textbf{0.0099}\rlap{$^*$} \\        
        \textbf{CaseRec-S} & {0.0066}\rlap{$^*$} & {0.0133}\rlap{$^*$} & {0.0224}\rlap{$^*$} & {0.0042}\rlap{$^*$} & {0.0063}\rlap{$^*$} & {0.0086}\rlap{$^*$} \\  
        \bottomrule
        \end{tabular}
    }
    \label{tab:debias_results}
\end{table}

\subsubsection{Datasets}
Experiments are conducted on three public-accessible datasets: ZhihuRec \cite{hao2021largescalezhihu}, Tenrec \cite{yuan2022tenrec} and KuaiRand \cite{gao2022kuairand}. All three datasets contain user interaction data (e.g., clicks) and system exposure data (e.g., impressions). We treat the clicked items as interacted items and non-clicked items as exposed but non-interacted items. 

\begin{itemize}[leftmargin=10pt]
    \item \textbf{ZhihuRec}\footnote{\url{https://github.com/THUIR/ZhihuRec-Dataset}} contains question information, answer information and use profiles. We consider the answers as the items recommended to the user.
    \item \textbf{Tenrec}\footnote{\url{https://github.com/yuangh-x/2022-NIPS-Tenrec}} is collected from two feeds, namely articles and videos, on Tenrec’s recommendation platforms. Our study focuses on the video recommendation scenario.
    \item \textbf{KuaiRand-Pure}\footnote{\url{https://kuairand.com/}} is a sequential recommendation dataset collected from video recommendation scenario. It contains both the standard user feedback records collected by prior recommendation policies (\textbf{KuaiRand-15policies}) and unbiased records collected by randomly exposing items (\textbf{KuaiRand-Random}).
\end{itemize}
\subsubsection{Evaluation protocols}
We employ Recall@$K$ and NDCG@$K$ as the evaluation metrics to evaluate the ranking performance. 
The results are reported with varying values of $K \in \{5,10,20\}$ for both metrics. 
We also adopt cross-validation to evaluate the performance of the models and the ratio of training, validation, and test set is 8:1:1.
To ensure a fair comparison, we use user interaction sequences as the input of inference, which is consistent with the baselines.
Each experiment is repeated 5 times, and the average performance is reported.

\subsubsection{Baselines} We compare CaseRec with various state-of-the-art sequential recommendation models, including RNN-based models, transformer-based models, debiasing models and RL-based models:
\begin{itemize}[leftmargin=10pt]
\item \textbf{GRU4Rec}\cite{DBLP:journals/corr/HidasiKBT15} is an RNN-based sequential recommendation model, which leverages GRU to model user–item interactions.
\item \textbf{SASRec}\cite{SASRec} leverages a left-to-right unidirectional Transformer to capture user preference.
\item \textbf{BERT4Rec}\cite{sun2019bert4rec} uses a bidirectional Transformer to learn sequential information.
\item \textbf{CORE}\cite{hou2022core} leverages a transformer-based structure to unify representation spaces for both the encoding and decoding processes.
\item \textbf{FEARec}\cite{du2023frequencyFEA} utilizes contrastive learning to capture hidden information within the frequency domain.
\item \textbf{IPS}\cite{ips_rec} eliminates popularity bias by re-weighing each interaction according to propensity score.
\item \textbf{RelMF}\cite{Yang2018unbiasevaluatuion} aims to use an effective unbiased estimator to correct the matching score between items and users.
\item \textbf{DRO}\cite{yang2024debiasing} utilizes the whole system exposure to alleviate the exposure bias by distributionally robust optimization (DRO). 
\item \textbf{DT}\cite{chen2021decision} is a vanilla DT model without considering the system exposure sequence. 
\item \textbf{SQN}\cite{xin2020self} is an RL-based model incorporating Q-learning into a standard supervised generative sequential model.
\end{itemize}

\subsubsection{Implementation details} 
Following previous work\cite{yang2024debiasing}, we preserve the last 50 clicked items as the user-interaction sequence and the last 200 exposed items as the system exposure sequence. 
The sequences are padded to the max length with an additional token. 
For CaseRec, we set the maximum interaction steps of trajectory as $T=20$, the maximum interaction steps of the state as $L=10$, the augmentation length as $h=10$, the Transformer layer number as 2, and the heads number as 8 for both the sequential recommender and user simulator.
For a fair comparison, the item embedding size is set to 64 for all models. We train all models with the Adam optimizer\cite{DBLP:journals/corr/KingmaB14}. 
For other baselines, the hyper-parameters are set to their optimal values as recommended by their original papers.

\begin{figure*}[t]
\centering
\includegraphics[width=1\linewidth]{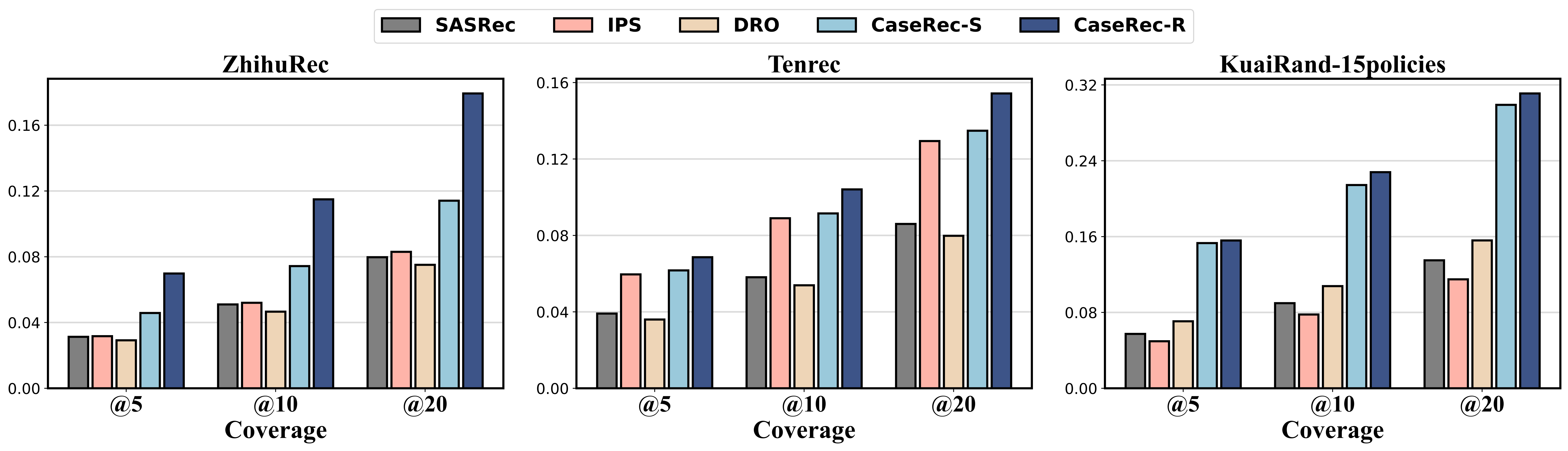}
\caption{Evaluation on recommendation diversity.}
\label{fig:exp_coverage}
\end{figure*}

\begin{figure}[t]
\centering
\includegraphics[width=0.97\linewidth]{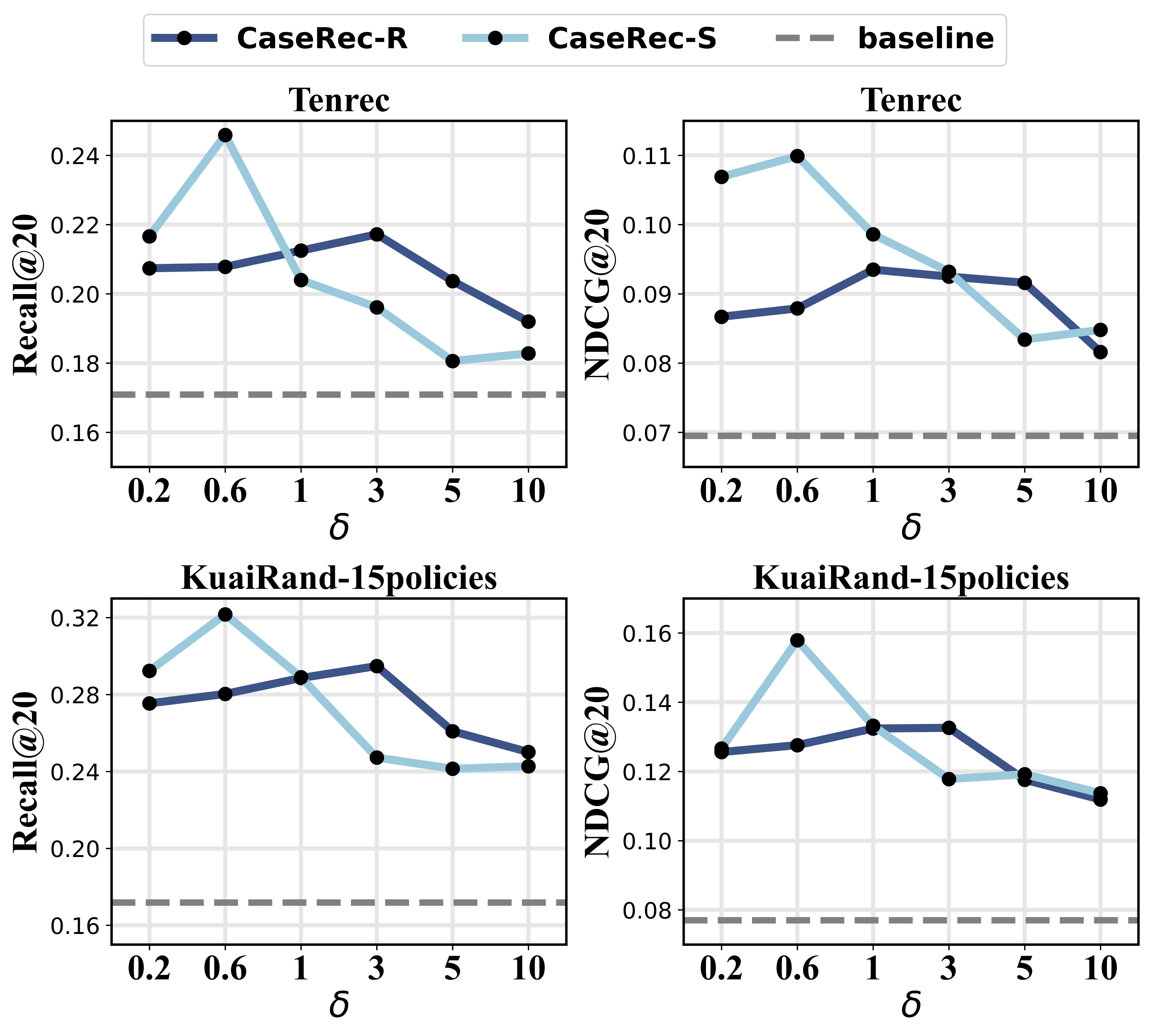}
\caption{Impact of the augmentation ratio $\delta$. Grey dashed lines denote the best baseline.}
\label{fig:exp_ratio}
\vspace{-5pt}
\end{figure}

\subsection{Overall performance (RQ1)}

Table \ref{tab:main_results} presents the overall performance compared with the baselines in three datasets. 
From Table \ref{tab:main_results}, we can observe that both CaseRec-S and CaseRec-R outperform all baselines across all datasets, demonstrating the effectiveness of our approach in modeling the system exposure.
Compared to existing methods only utilizing the sparse user-interaction sequence, our method takes full advantage of the large volume of exposure sequence to capture the potential user interests.
DRO\cite{yang2024debiasing} also utilized the informative system exposure to model the system behavior and achieve improved performance, particularly in complex datasets with large item space like Tenrec. 
However, our method still shows a significant performance improvement compared to DRO, which demonstrates that the counterfactual data generated by the data augmenter helps the sequential recommender to further uncover the user interests. 

Additionally, we can observe that CaseRec-S outperforms CaseRec-R, which may be attributed to the reason that CaseRec-S explores the item space near the historical policy while CaseRec-R imitates complete random exploration.
Finally, we observe that the improvement of CaseRec is most significant on the KuaiRand-15policies dataset, which has a higher density than other datasets. 
We assume that the user simulator can more accurately capture user interests from the dense information and lead to higher-quality generated sequence augmentation.

\subsection{Debiasing performance (RQ2)}
In this section, we conduct experiments to compare the debiasing performance of CaseRec with other debiasing methods. 

\textbf{Evaluation on unbiased exposure.}
Exposure bias, also known as ``previous model bias'' \cite{liu2020general}, occurs when the recommendation model is influenced by prior recommendation policies and only exposes a subset of available items to users. 
Table \ref{tab:debias_results} presents the debiasing performance in the unbiased KuaiRand-Random dataset, which is collected by exposing items to users randomly rather than by prior recommendation policies.
Specifically, to evaluate the unbiased recommendation performance learned from the biased data, we use the biased KuaiRand-15policies dataset as the training set and the unbiased KuaiRand-Random dataset as the validation and test set for each method.

From Table \ref{tab:debias_results}, we see that only CaseRec achieves satisfying performance in this challenging experimental setting, 
which indicates that the counterfactual augmentation captures new user interests beyond the original dataset collected using the previous recommendation policies.
Existing debiasing methods (IPS, RelMF and DRO) evaluate items and introduce penalties on those considered overexposed, achieving improved performance compared to the vanilla models. 
Evaluating whether an item is overexposed is a challenging task and imposing penalties on inappropriate items negatively impacts the model performance.
CaseRec-R outperforms CaseRec-S in this setting, which indicates that the exposure sequence generated by random expose simulation helps to reduce the impact of prior policies.

\textbf{Evaluation on recommendation diversity.} 
Exposure bias could also cause relevant but unpopular items to be overlooked and not recommended, leading to a reduction in recommendation diversity~\cite{biassurvey,gupta2021correcting}.
We conduct experiments to determine whether CaseRec can generate diversified recommendation.
We use the coverage metric to measure the recommendation diversity:
\begin{equation}
    \text{Coverage@}K=\frac{|\bigcup_{u \in \text{Test}} list_{@K}(u)|}{|\mathcal{V}|}
\end{equation}
where $ list_{@K}(u)$ represents the the top-$K$ recommendation list for user $u$.
Figure \ref{fig:exp_coverage} presents the exposure debiasing performance compared with baselines in three datasets. 
From Figure \ref{fig:exp_coverage}, we can observe that CaseRec-R achieves the best performance across all datasets, which indicates that the baselines only depend on the logged historical sequence, which only contains a small subset of all available items. 
In contrast, our approach explores user interests over a broader range of items, leading to more diverse lists.

To conclude, our method can not only achieve better recommendation performance under unbiased evaluation but also generate more diverse recommendation. 
This suggests that our method is an alternative solution to alleviate the exposure bias.

\subsection{Impact of augmentation ratio (RQ3)}
In this section, we conduct experiments to demonstrate the impact of the augmentation ratio $\delta$, which can be defined as $\delta=\frac{|\mathcal{D}_{aug}|}{|\mathcal{D}_{ori}|}$, where ${D}_{aug}$ represents the augmented dataset and ${D}_{ori}$ represents the original dataset. 
Specifically, we set the $\delta$ varying from 0.2 to 10 and evaluate the performance for each augmentation ratio. 
The results are presented in Figure \ref{fig:exp_ratio}. 
We can observe that the performance of both CaseRec-R and CaseRec-S in both datasets first increases and then decreases. 
This indicates that when the augmentation ratio increases at the beginning, the performance of CaseRec improves because the augmented exposure sequence enables the sequential recommender to further explore the user interests.
However, when the augmentation ratio is too large, the performance will drop because too much noise could be introduced to the dataset.
In addition, compared to CaseRec-R, CaseRec-S is able to continuously improve the current recommendation strategy based on the user simulator, thus demonstrating higher learning efficiency.
Finally, regardless of the augmentation ratio, both CaseRec-S and  CaseRec-R outperform the strongest baseline, highlighting the robustness of our proposed method.

\subsection{Ablation study (RQ4)}
\begin{figure}[t]
\centering
\includegraphics[width=\linewidth]{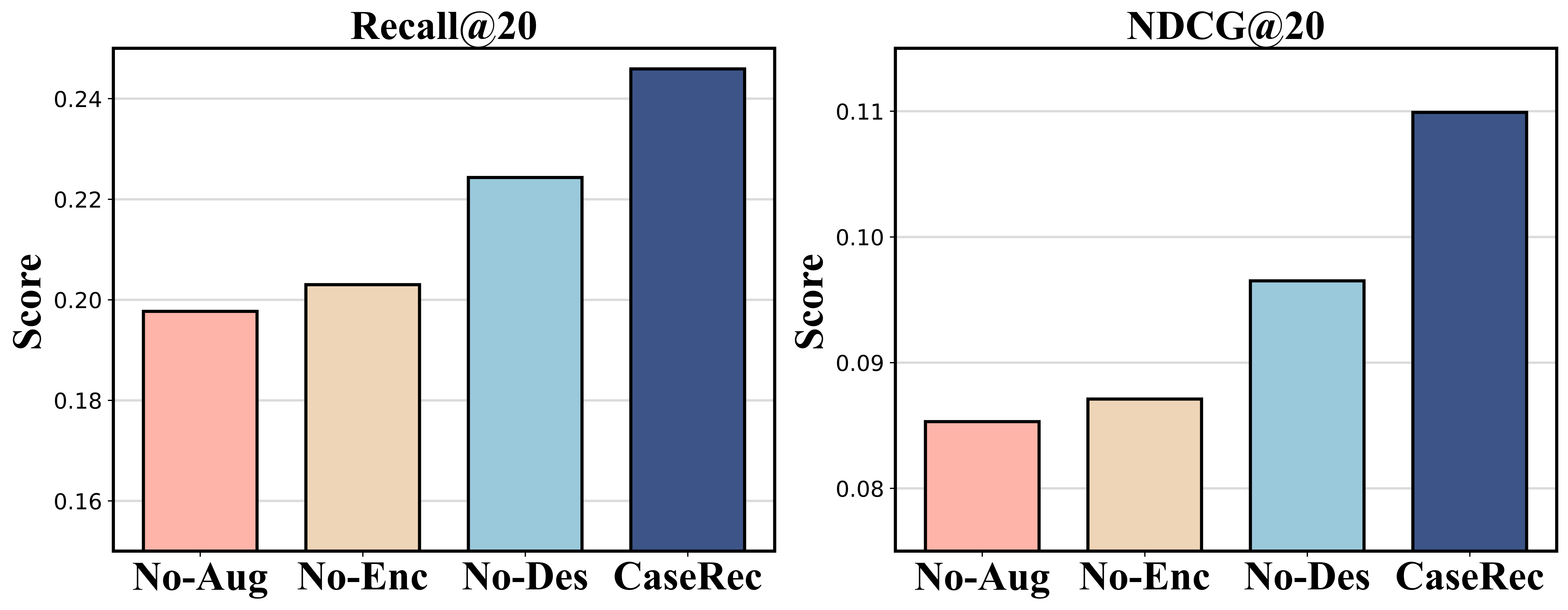}
\caption{Ablation Study.}
\label{fig:ablation}
\end{figure}

In this section, we perform an ablation study to investigate the impact of each module individually. Specifically, the ablation study was conducted using the Tenrec dataset based on the \textit{Self-Improving} strategy, with the results presented in Figure \ref{fig:ablation}. The variant models are as follows:
\begin{itemize}[leftmargin=10pt]
    \item \textbf{No-Aug}: The recommendation model is trained using only the original dataset, without the counterfactual augmentation.
    \item \textbf{No-Enc}: The recommendation model uses the average of the item embeddings in state sequence to replace the state encoder.
    \item \textbf{No-Des}: The recommendation model without the adaption of learning objectives.
\end{itemize}
From Figure \ref{fig:ablation}, we have several observations.
Firstly, CaseRec significantly outperforms 'No-Aug', demonstrating the effectiveness of the counterfactual augmentation. 
Besides, although 'No-Enc' shows an improvement compared to 'No-Aug', it still lags behind CaseRec, which demonstrates the effectiveness of the state encoder in simultaneously capturing sequential signals and user behaviors.
Finally, CaseRec outperforms 'No-Des', indicating that the redesigned learning objective effectively narrows the gap between original DT and SR, further improving performance.

\section{Related work}

In this section, we provide a literature review on sequential recommendation, offline reinforcement learning and system exposure. 
\subsection{Sequential recommendation}
The purpose of sequential recommendation (SR) is to capture the item’s chronological correlations. Early SR models are mainly based on the Markov Chain ~\cite{he2016vista,rendle2010factorizing}.
Over the past few years, many deep learning-based SR models have been proposed. 
For example, GRU4Rec~\cite{DBLP:journals/corr/HidasiKBT15} attempts to utilize recurrent neural networks to learn sequential information. 
Besides, the self-attention mechanism has shown great potential and various models have been proposed, e.g., SASRec ~\cite{SASRec}, BERT4Rec~\cite{sun2019bert4rec} and $S^{3}$Rec~\cite{zhou2020s3rec}. 
Furthermore, CORE ~\cite{hou2022core} proposes an effective transformer-based structure to unify the representation space for both the encoding and decoding processes.
Recently, many contrastive learning-based sequential recommendation methods have been proposed to further improve the performance. 
For example, FEARec~\cite{du2023frequencyFEA} separates low-frequency information and high-frequency information in the frequency domain and utilizes contrastive learning to capture hidden information within the frequency domain.

However, the aforementioned SR models are still based on traditional user interaction sequences and are unable to incorporate system exposure.
This prevents them from fully exploiting the user interests contained in system exposure data, leading to sub-optimal recommendation performance.

\subsection{Offline reinforcement learning}
Offline RL aims to learn policies from a pre-collected, static dataset of offline trajectories. 
Traditional offline RL methods can be classified into model-free methods and model-based methods.
Model-free offline methods always incorporate conservatism into the value function estimation~\cite{fujimoto2019off,kumar2019stabilizing,fujimoto2018addressing,siegel2020keep,kumar2020conservative,kostrikov2021offlinefisher}.
For example, CQL~\cite{kumar2020conservative} adds a regularization term into the Q-function update.
Model-based offline RL methods~\cite{yu2020mopo,kidambi2020morel,yu2021combo,rigter2022rambo, zhan2022modelbased, swazinna2021overcoming, rafailov2020offline, matsushima2020deploymentefficient,lowrey2018plan,schrittwieser2021online} learn a model of environment and generate additional data to improve the policy.
Unlike the aforementioned two types of methods, Decision Transformer~\cite{chen2021decision} casts the offline RL task as a conditional sequence modeling problem and achieves remarkable performance.

Meanwhile, offline RL has also achieved successful performance in sequential recommendation scenario~\cite{chen2023deepsurvey,afsar2022reinforcement,wang2022surrogate,xin2020self,zhao2018deep}. 
For example, SQN~\cite{xin2020self} incorporates an additional output layer trained with Q-learning into a standard supervised generative sequential model and achieves improved performance.
Recently, some DT-based SR models have been proposed~\cite{zhou2020s3rec,liu2024sequential}. However, unlike our work, these models are user retention-oriented by treating long-term user engagement as an additional learning objective of DT.
On the contrary, our work aims to generate more accurate recommendation by leveraging the characteristics of DT to model the complex system exposure sequences.


\subsection{System exposure}
System exposure data, also referred to as \textit{previous recommendation} \cite{azzalini2022socrate} or \textit{impressions} \cite{maurera2023impression}, represents the system's behavior as opposed to user interaction data, which reflects user preferences. Despite its relevance, system exposure data has received relatively little attention in the research community. 
Existing system exposure studies primarily focus on the negative sampling of exposed items or addressing exposure bias.

Negative sampling is a technique in which a subset of unobserved or non-preferred data points is randomly selected and treated as negative examples to improve model training efficiency. 
\citet{ding2018improved} propose leveraging the exposed but not interacted items as negative examples. 
\citet{wang2020reinforced} propose a reinforcement negative sampler that generates exposure-like negative instances rather than directly selecting from the exposure data.
Instead of simply considering the exposed but non-interacted items as negative examples, we argue that exposed but non-interacted items could also contain potential user interests.

Exposure bias happens when only parts of specific items are exposed by the system, so it is unclear whether an unobserved interaction between a user and an item is attributable to the user's lack of interest or the item was not exposed to the user.
Therefore, regarding non-interacted items as negative samples could lead to a misunderstanding of user's true preference and sub-optimal performance.
Existing exposure debiasing methods typically introduce penalties based on IPS or DRO.
For instance, \citet{Yang2018unbiasevaluatuion} proposes an IPS-based unbiased evaluator to down-weight the commonly observed interactions while up-weighting the rare ones.
\citet{yang2024debiasing} uses DRO to infer the debiased user preference and introduces penalties to items with high exposure probability.
Our method introduces counterfactual exposure sequences and can be considered as an alternative solution to further explore user interests, leading to downgraded exposure bias.

\section{Conclusion}

In this paper, we have presented a novel framework, i.e., CaseRec, to further enhance the sequential recommendation.
CaseRec uses a DT-based sequential model to take the whole exposure sequence as input to generate recommendation. 
In addition, CaseRec performs two counterfactual augmentation strategies, namely strategy \textit{Random} and strategy \textit{Self-Improving}, to further uncover potential user interests.
Then, a transformer-based user simulator has been proposed to predict the user's corresponding feedback towards the augmented items.
Extensive experiments on three real-world datasets have demonstrated the effectiveness of the proposed CaseRec. 

In this paper, we have considered only two types of user behaviors. In future work, we plan to incorporate a broader range of user behaviors.
Moreover, there is a rapidly growing body of work on user simulators~\cite{balog-2024-user}. In future work, we aim to experiment with a broader set of user simulators.

\clearpage
\newpage

\bibliographystyle{ACM-Reference-Format}
\bibliography{references}


\begin{thebibliography}{63}


\ifx \showCODEN    \undefined \def \showCODEN     #1{\unskip}     \fi
\ifx \showDOI      \undefined \def \showDOI       #1{#1}\fi
\ifx \showISBNx    \undefined \def \showISBNx     #1{\unskip}     \fi
\ifx \showISBNxiii \undefined \def \showISBNxiii  #1{\unskip}     \fi
\ifx \showISSN     \undefined \def \showISSN      #1{\unskip}     \fi
\ifx \showLCCN     \undefined \def \showLCCN      #1{\unskip}     \fi
\ifx \shownote     \undefined \def \shownote      #1{#1}          \fi
\ifx \showarticletitle \undefined \def \showarticletitle #1{#1}   \fi
\ifx \showURL      \undefined \def \showURL       {\relax}        \fi
\providecommand\bibfield[2]{#2}
\providecommand\bibinfo[2]{#2}
\providecommand\natexlab[1]{#1}
\providecommand\showeprint[2][]{arXiv:#2}

\bibitem[Afsar et~al\mbox{.}(2022)]%
        {afsar2022reinforcement}
\bibfield{author}{\bibinfo{person}{M~Mehdi Afsar}, \bibinfo{person}{Trafford Crump}, {and} \bibinfo{person}{Behrouz Far}.} \bibinfo{year}{2022}\natexlab{}.
\newblock \showarticletitle{Reinforcement learning based recommender systems: A survey}.
\newblock \bibinfo{journal}{\emph{Comput. Surveys}} \bibinfo{volume}{55}, \bibinfo{number}{7} (\bibinfo{year}{2022}), \bibinfo{pages}{1--38}.
\newblock


\bibitem[Azzalini et~al\mbox{.}(2022)]%
        {azzalini2022socrate}
\bibfield{author}{\bibinfo{person}{Davide Azzalini}, \bibinfo{person}{Fabio Azzalini}, \bibinfo{person}{Chiara Criscuolo}, \bibinfo{person}{Tommaso Dolci}, \bibinfo{person}{Davide Martinenghi}, {and} \bibinfo{person}{Sihem Amer-Yahia}.} \bibinfo{year}{2022}\natexlab{}.
\newblock \showarticletitle{SoCRATe: A recommendation system with limited-availability items}. In \bibinfo{booktitle}{\emph{Proceedings of the 31st ACM International Conference on Information \& Knowledge Management}}. \bibinfo{pages}{4793--4797}.
\newblock


\bibitem[Ba(2016)]%
        {ba2016layer}
\bibfield{author}{\bibinfo{person}{Jimmy~Lei Ba}.} \bibinfo{year}{2016}\natexlab{}.
\newblock \showarticletitle{Layer normalization}.
\newblock \bibinfo{journal}{\emph{arXiv preprint arXiv:1607.06450}} (\bibinfo{year}{2016}).
\newblock


\bibitem[Balog and Zhai(2024)]%
        {balog-2024-user}
\bibfield{author}{\bibinfo{person}{Krisztian Balog} {and} \bibinfo{person}{ChengXiang Zhai}.} \bibinfo{year}{2024}\natexlab{}.
\newblock \showarticletitle{User Simulation for Evaluating Information Access Systems}.
\newblock \bibinfo{journal}{\emph{Foundations and Trends{\textregistered} in Information Retrieval}} \bibinfo{volume}{18}, \bibinfo{number}{1-2} (\bibinfo{year}{2024}), \bibinfo{pages}{1--261}.
\newblock


\bibitem[Chen et~al\mbox{.}(2023a)]%
        {biassurvey}
\bibfield{author}{\bibinfo{person}{Jiawei Chen}, \bibinfo{person}{Hande Dong}, \bibinfo{person}{Xiang Wang}, \bibinfo{person}{Fuli Feng}, \bibinfo{person}{Meng Wang}, {and} \bibinfo{person}{Xiangnan He}.} \bibinfo{year}{2023}\natexlab{a}.
\newblock \showarticletitle{Bias and debias in recommender system: A survey and future directions}.
\newblock \bibinfo{journal}{\emph{ACM Transactions on Information Systems}} \bibinfo{volume}{41}, \bibinfo{number}{3} (\bibinfo{year}{2023}), \bibinfo{pages}{1--39}.
\newblock


\bibitem[Chen et~al\mbox{.}(2021)]%
        {chen2021decision}
\bibfield{author}{\bibinfo{person}{Lili Chen}, \bibinfo{person}{Kevin Lu}, \bibinfo{person}{Aravind Rajeswaran}, \bibinfo{person}{Kimin Lee}, \bibinfo{person}{Aditya Grover}, \bibinfo{person}{Misha Laskin}, \bibinfo{person}{Pieter Abbeel}, \bibinfo{person}{Aravind Srinivas}, {and} \bibinfo{person}{Igor Mordatch}.} \bibinfo{year}{2021}\natexlab{}.
\newblock \showarticletitle{Decision transformer: Reinforcement learning via sequence modeling}.
\newblock \bibinfo{journal}{\emph{Advances in neural information processing systems}}  \bibinfo{volume}{34} (\bibinfo{year}{2021}), \bibinfo{pages}{15084--15097}.
\newblock


\bibitem[Chen et~al\mbox{.}(2023b)]%
        {chen2023deepsurvey}
\bibfield{author}{\bibinfo{person}{Xiaocong Chen}, \bibinfo{person}{Lina Yao}, \bibinfo{person}{Julian McAuley}, \bibinfo{person}{Guanglin Zhou}, {and} \bibinfo{person}{Xianzhi Wang}.} \bibinfo{year}{2023}\natexlab{b}.
\newblock \showarticletitle{Deep reinforcement learning in recommender systems: A survey and new perspectives}.
\newblock \bibinfo{journal}{\emph{Knowledge-Based Systems}}  \bibinfo{volume}{264} (\bibinfo{year}{2023}), \bibinfo{pages}{110335}.
\newblock


\bibitem[Cho(2014)]%
        {cho2014learning}
\bibfield{author}{\bibinfo{person}{Kyunghyun Cho}.} \bibinfo{year}{2014}\natexlab{}.
\newblock \showarticletitle{Learning phrase representations using RNN encoder-decoder for statistical machine translation}.
\newblock \bibinfo{journal}{\emph{arXiv preprint arXiv:1406.1078}} (\bibinfo{year}{2014}).
\newblock


\bibitem[Dai et~al\mbox{.}(2022)]%
        {dai2022generalized}
\bibfield{author}{\bibinfo{person}{Quanyu Dai}, \bibinfo{person}{Haoxuan Li}, \bibinfo{person}{Peng Wu}, \bibinfo{person}{Zhenhua Dong}, \bibinfo{person}{Xiao-Hua Zhou}, \bibinfo{person}{Rui Zhang}, \bibinfo{person}{Rui Zhang}, {and} \bibinfo{person}{Jie Sun}.} \bibinfo{year}{2022}\natexlab{}.
\newblock \showarticletitle{A generalized doubly robust learning framework for debiasing post-click conversion rate prediction}. In \bibinfo{booktitle}{\emph{KDD}}. \bibinfo{pages}{252--262}.
\newblock


\bibitem[Damak et~al\mbox{.}(2022)]%
        {Khalil2022cloze}
\bibfield{author}{\bibinfo{person}{Khalil Damak}, \bibinfo{person}{Sami Khenissi}, {and} \bibinfo{person}{Olfa Nasraoui}.} \bibinfo{year}{2022}\natexlab{}.
\newblock \showarticletitle{Debiasing the Cloze Task in Sequential Recommendation with Bidirectional Transformers}. In \bibinfo{booktitle}{\emph{KDD}}. \bibinfo{pages}{273--282}.
\newblock


\bibitem[Ding et~al\mbox{.}(2018)]%
        {ding2018improved}
\bibfield{author}{\bibinfo{person}{Jingtao Ding}, \bibinfo{person}{Fuli Feng}, \bibinfo{person}{Xiangnan He}, \bibinfo{person}{Guanghui Yu}, \bibinfo{person}{Yong Li}, {and} \bibinfo{person}{Depeng Jin}.} \bibinfo{year}{2018}\natexlab{}.
\newblock \showarticletitle{An improved sampler for bayesian personalized ranking by leveraging view data}. In \bibinfo{booktitle}{\emph{Companion Proceedings of the The Web Conference 2018}}. \bibinfo{pages}{13--14}.
\newblock


\bibitem[Du et~al\mbox{.}(2023)]%
        {du2023frequencyFEA}
\bibfield{author}{\bibinfo{person}{Xinyu Du}, \bibinfo{person}{Huanhuan Yuan}, \bibinfo{person}{Pengpeng Zhao}, \bibinfo{person}{Jianfeng Qu}, \bibinfo{person}{Fuzhen Zhuang}, \bibinfo{person}{Guanfeng Liu}, \bibinfo{person}{Yanchi Liu}, {and} \bibinfo{person}{Victor~S Sheng}.} \bibinfo{year}{2023}\natexlab{}.
\newblock \showarticletitle{Frequency enhanced hybrid attention network for sequential recommendation}. In \bibinfo{booktitle}{\emph{Proceedings of the 46th International ACM SIGIR Conference on Research and Development in Information Retrieval}}. \bibinfo{pages}{78--88}.
\newblock


\bibitem[Fujimoto et~al\mbox{.}(2018)]%
        {fujimoto2018addressing}
\bibfield{author}{\bibinfo{person}{Scott Fujimoto}, \bibinfo{person}{Herke Hoof}, {and} \bibinfo{person}{David Meger}.} \bibinfo{year}{2018}\natexlab{}.
\newblock \showarticletitle{Addressing function approximation error in actor-critic methods}. In \bibinfo{booktitle}{\emph{International conference on machine learning}}. PMLR, \bibinfo{pages}{1587--1596}.
\newblock


\bibitem[Fujimoto et~al\mbox{.}(2019)]%
        {fujimoto2019off}
\bibfield{author}{\bibinfo{person}{Scott Fujimoto}, \bibinfo{person}{David Meger}, {and} \bibinfo{person}{Doina Precup}.} \bibinfo{year}{2019}\natexlab{}.
\newblock \showarticletitle{Off-policy deep reinforcement learning without exploration}. In \bibinfo{booktitle}{\emph{International conference on machine learning}}. PMLR, \bibinfo{pages}{2052--2062}.
\newblock


\bibitem[Gao et~al\mbox{.}(2022)]%
        {gao2022kuairand}
\bibfield{author}{\bibinfo{person}{Chongming Gao}, \bibinfo{person}{Shijun Li}, \bibinfo{person}{Yuan Zhang}, \bibinfo{person}{Jiawei Chen}, \bibinfo{person}{Biao Li}, \bibinfo{person}{Wenqiang Lei}, \bibinfo{person}{Peng Jiang}, {and} \bibinfo{person}{Xiangnan He}.} \bibinfo{year}{2022}\natexlab{}.
\newblock \showarticletitle{KuaiRand: An Unbiased Sequential Recommendation Dataset with Randomly Exposed Videos}. In \bibinfo{booktitle}{\emph{Proceedings of the 31st ACM International Conference on Information and Knowledge Management}} (Atlanta, GA, USA) \emph{(\bibinfo{series}{CIKM '22})}. \bibinfo{pages}{3953–3957}.
\newblock
\urldef\tempurl%
\url{https://doi.org/10.1145/3511808.3557624}
\showDOI{\tempurl}


\bibitem[Gupta et~al\mbox{.}(2021)]%
        {gupta2021correcting}
\bibfield{author}{\bibinfo{person}{Shantanu Gupta}, \bibinfo{person}{Hao Wang}, \bibinfo{person}{Zachary Lipton}, {and} \bibinfo{person}{Yuyang Wang}.} \bibinfo{year}{2021}\natexlab{}.
\newblock \showarticletitle{Correcting exposure bias for link recommendation}. In \bibinfo{booktitle}{\emph{International Conference on Machine Learning}}. PMLR, \bibinfo{pages}{3953--3963}.
\newblock


\bibitem[Hao et~al\mbox{.}(2021)]%
        {hao2021largescalezhihu}
\bibfield{author}{\bibinfo{person}{Bin Hao}, \bibinfo{person}{Min Zhang}, \bibinfo{person}{Weizhi Ma}, \bibinfo{person}{Shaoyun Shi}, \bibinfo{person}{Xinxing Yu}, \bibinfo{person}{Houzhi Shan}, \bibinfo{person}{Yiqun Liu}, {and} \bibinfo{person}{Shaoping Ma}.} \bibinfo{year}{2021}\natexlab{}.
\newblock \bibinfo{title}{A Large-Scale Rich Context Query and Recommendation Dataset in Online Knowledge-Sharing}.
\newblock
\newblock
\showeprint[arxiv]{2106.06467}~[cs.IR]


\bibitem[He et~al\mbox{.}(2016)]%
        {he2016vista}
\bibfield{author}{\bibinfo{person}{Ruining He}, \bibinfo{person}{Chen Fang}, \bibinfo{person}{Zhaowen Wang}, {and} \bibinfo{person}{Julian McAuley}.} \bibinfo{year}{2016}\natexlab{}.
\newblock \showarticletitle{Vista: A visually, socially, and temporally-aware model for artistic recommendation}. In \bibinfo{booktitle}{\emph{Proceedings of the 10th ACM conference on recommender systems}}. \bibinfo{pages}{309--316}.
\newblock


\bibitem[He et~al\mbox{.}(2017)]%
        {he2017ncf}
\bibfield{author}{\bibinfo{person}{Xiangnan He}, \bibinfo{person}{Lizi Liao}, \bibinfo{person}{Hanwang Zhang}, \bibinfo{person}{Liqiang Nie}, \bibinfo{person}{Xia Hu}, {and} \bibinfo{person}{Tat-Seng Chua}.} \bibinfo{year}{2017}\natexlab{}.
\newblock \showarticletitle{Neural collaborative filtering}. In \bibinfo{booktitle}{\emph{WWW}}. \bibinfo{pages}{173--182}.
\newblock


\bibitem[Hendrycks and Gimpel(2016)]%
        {hendrycks2016gaussian}
\bibfield{author}{\bibinfo{person}{Dan Hendrycks} {and} \bibinfo{person}{Kevin Gimpel}.} \bibinfo{year}{2016}\natexlab{}.
\newblock \showarticletitle{Gaussian error linear units (gelus)}.
\newblock \bibinfo{journal}{\emph{arXiv preprint arXiv:1606.08415}} (\bibinfo{year}{2016}).
\newblock


\bibitem[Hidasi et~al\mbox{.}(2016)]%
        {DBLP:journals/corr/HidasiKBT15}
\bibfield{author}{\bibinfo{person}{Bal{\'{a}}zs Hidasi}, \bibinfo{person}{Alexandros Karatzoglou}, \bibinfo{person}{Linas Baltrunas}, {and} \bibinfo{person}{Domonkos Tikk}.} \bibinfo{year}{2016}\natexlab{}.
\newblock \showarticletitle{Session-based Recommendations with Recurrent Neural Networks}. In \bibinfo{booktitle}{\emph{{ICLR} (Poster)}}.
\newblock


\bibitem[Hou et~al\mbox{.}(2022)]%
        {hou2022core}
\bibfield{author}{\bibinfo{person}{Yupeng Hou}, \bibinfo{person}{Binbin Hu}, \bibinfo{person}{Zhiqiang Zhang}, {and} \bibinfo{person}{Wayne~Xin Zhao}.} \bibinfo{year}{2022}\natexlab{}.
\newblock \showarticletitle{Core: simple and effective session-based recommendation within consistent representation space}. In \bibinfo{booktitle}{\emph{Proceedings of the 45th international ACM SIGIR conference on research and development in information retrieval}}. \bibinfo{pages}{1796--1801}.
\newblock


\bibitem[Hu et~al\mbox{.}(2018)]%
        {reinforce-e-commerce}
\bibfield{author}{\bibinfo{person}{Yujing Hu}, \bibinfo{person}{Qing Da}, \bibinfo{person}{Anxiang Zeng}, \bibinfo{person}{Yang Yu}, {and} \bibinfo{person}{Yinghui Xu}.} \bibinfo{year}{2018}\natexlab{}.
\newblock \showarticletitle{Reinforcement learning to rank in e-commerce search engine: Formalization, analysis, and application}. In \bibinfo{booktitle}{\emph{KDD}}. ACM, \bibinfo{pages}{368--377}.
\newblock


\bibitem[Jiang et~al\mbox{.}(2019)]%
        {10.1145/3306618.3314288}
\bibfield{author}{\bibinfo{person}{Ray Jiang}, \bibinfo{person}{Silvia Chiappa}, \bibinfo{person}{Tor Lattimore}, \bibinfo{person}{Andr{\'a}s Gy{\"o}rgy}, {and} \bibinfo{person}{Pushmeet Kohli}.} \bibinfo{year}{2019}\natexlab{}.
\newblock \showarticletitle{Degenerate feedback loops in recommender systems}. In \bibinfo{booktitle}{\emph{Proceedings of the 2019 AAAI/ACM Conference on AI, Ethics, and Society}}. \bibinfo{pages}{383--390}.
\newblock


\bibitem[Kang and McAuley(2018)]%
        {SASRec}
\bibfield{author}{\bibinfo{person}{Wang-Cheng Kang} {and} \bibinfo{person}{Julian McAuley}.} \bibinfo{year}{2018}\natexlab{}.
\newblock \showarticletitle{Self-attentive sequential recommendation}. In \bibinfo{booktitle}{\emph{ICDM}}. IEEE, \bibinfo{pages}{197--206}.
\newblock


\bibitem[Kidambi et~al\mbox{.}(2020)]%
        {kidambi2020morel}
\bibfield{author}{\bibinfo{person}{Rahul Kidambi}, \bibinfo{person}{Aravind Rajeswaran}, \bibinfo{person}{Praneeth Netrapalli}, {and} \bibinfo{person}{Thorsten Joachims}.} \bibinfo{year}{2020}\natexlab{}.
\newblock \showarticletitle{Morel: Model-based offline reinforcement learning}.
\newblock \bibinfo{journal}{\emph{Advances in neural information processing systems}}  \bibinfo{volume}{33} (\bibinfo{year}{2020}), \bibinfo{pages}{21810--21823}.
\newblock


\bibitem[Kingma and Ba(2015)]%
        {DBLP:journals/corr/KingmaB14}
\bibfield{author}{\bibinfo{person}{Diederik~P. Kingma} {and} \bibinfo{person}{Jimmy Ba}.} \bibinfo{year}{2015}\natexlab{}.
\newblock \showarticletitle{Adam: {A} Method for Stochastic Optimization}. In \bibinfo{booktitle}{\emph{{ICLR} (Poster)}}.
\newblock


\bibitem[Kostrikov et~al\mbox{.}(2021)]%
        {kostrikov2021offlinefisher}
\bibfield{author}{\bibinfo{person}{Ilya Kostrikov}, \bibinfo{person}{Rob Fergus}, \bibinfo{person}{Jonathan Tompson}, {and} \bibinfo{person}{Ofir Nachum}.} \bibinfo{year}{2021}\natexlab{}.
\newblock \showarticletitle{Offline reinforcement learning with fisher divergence critic regularization}. In \bibinfo{booktitle}{\emph{International Conference on Machine Learning}}. PMLR, \bibinfo{pages}{5774--5783}.
\newblock


\bibitem[Kumar et~al\mbox{.}(2019)]%
        {kumar2019stabilizing}
\bibfield{author}{\bibinfo{person}{Aviral Kumar}, \bibinfo{person}{Justin Fu}, \bibinfo{person}{Matthew Soh}, \bibinfo{person}{George Tucker}, {and} \bibinfo{person}{Sergey Levine}.} \bibinfo{year}{2019}\natexlab{}.
\newblock \showarticletitle{Stabilizing off-policy q-learning via bootstrapping error reduction}.
\newblock \bibinfo{journal}{\emph{Advances in Neural Information Processing Systems}}  \bibinfo{volume}{32} (\bibinfo{year}{2019}).
\newblock


\bibitem[Kumar et~al\mbox{.}(2020)]%
        {kumar2020conservative}
\bibfield{author}{\bibinfo{person}{Aviral Kumar}, \bibinfo{person}{Aurick Zhou}, \bibinfo{person}{George Tucker}, {and} \bibinfo{person}{Sergey Levine}.} \bibinfo{year}{2020}\natexlab{}.
\newblock \showarticletitle{Conservative q-learning for offline reinforcement learning}.
\newblock \bibinfo{journal}{\emph{Advances in Neural Information Processing Systems}}  \bibinfo{volume}{33} (\bibinfo{year}{2020}), \bibinfo{pages}{1179--1191}.
\newblock


\bibitem[Liang et~al\mbox{.}(2016)]%
        {DBLP:conf/www/expomf}
\bibfield{author}{\bibinfo{person}{Dawen Liang}, \bibinfo{person}{Laurent Charlin}, \bibinfo{person}{James McInerney}, {and} \bibinfo{person}{David~M Blei}.} \bibinfo{year}{2016}\natexlab{}.
\newblock \showarticletitle{Modeling user exposure in recommendation}. In \bibinfo{booktitle}{\emph{www}}. \bibinfo{pages}{951--961}.
\newblock


\bibitem[Liu et~al\mbox{.}(2020)]%
        {liu2020general}
\bibfield{author}{\bibinfo{person}{Dugang Liu}, \bibinfo{person}{Pengxiang Cheng}, \bibinfo{person}{Zhenhua Dong}, \bibinfo{person}{Xiuqiang He}, \bibinfo{person}{Weike Pan}, {and} \bibinfo{person}{Zhong Ming}.} \bibinfo{year}{2020}\natexlab{}.
\newblock \showarticletitle{A general knowledge distillation framework for counterfactual recommendation via uniform data}. In \bibinfo{booktitle}{\emph{Proceedings of the 43rd international ACM SIGIR conference on research and development in information retrieval}}. \bibinfo{pages}{831--840}.
\newblock


\bibitem[Liu et~al\mbox{.}(2023)]%
        {liu2023bounding}
\bibfield{author}{\bibinfo{person}{Dugang Liu}, \bibinfo{person}{Pengxiang Cheng}, \bibinfo{person}{Zinan Lin}, \bibinfo{person}{Xiaolian Zhang}, \bibinfo{person}{Zhenhua Dong}, \bibinfo{person}{Rui Zhang}, \bibinfo{person}{Xiuqiang He}, \bibinfo{person}{Weike Pan}, {and} \bibinfo{person}{Zhong Ming}.} \bibinfo{year}{2023}\natexlab{}.
\newblock \showarticletitle{Bounding System-Induced Biases in Recommender Systems with a Randomized Dataset}.
\newblock \bibinfo{journal}{\emph{ACM Transactions on Information Systems}} \bibinfo{volume}{41}, \bibinfo{number}{4} (\bibinfo{year}{2023}), \bibinfo{pages}{1--26}.
\newblock


\bibitem[Liu et~al\mbox{.}(2024)]%
        {liu2024sequential}
\bibfield{author}{\bibinfo{person}{Ziru Liu}, \bibinfo{person}{Shuchang Liu}, \bibinfo{person}{Zijian Zhang}, \bibinfo{person}{Qingpeng Cai}, \bibinfo{person}{Xiangyu Zhao}, \bibinfo{person}{Kesen Zhao}, \bibinfo{person}{Lantao Hu}, \bibinfo{person}{Peng Jiang}, {and} \bibinfo{person}{Kun Gai}.} \bibinfo{year}{2024}\natexlab{}.
\newblock \showarticletitle{Sequential Recommendation for Optimizing Both Immediate Feedback and Long-term Retention}. In \bibinfo{booktitle}{\emph{Proceedings of the 47th International ACM SIGIR Conference on Research and Development in Information Retrieval}}. \bibinfo{pages}{1872--1882}.
\newblock


\bibitem[Lowrey et~al\mbox{.}(2018)]%
        {lowrey2018plan}
\bibfield{author}{\bibinfo{person}{Kendall Lowrey}, \bibinfo{person}{Aravind Rajeswaran}, \bibinfo{person}{Sham Kakade}, \bibinfo{person}{Emanuel Todorov}, {and} \bibinfo{person}{Igor Mordatch}.} \bibinfo{year}{2018}\natexlab{}.
\newblock \showarticletitle{Plan online, learn offline: Efficient learning and exploration via model-based control}.
\newblock \bibinfo{journal}{\emph{arXiv preprint arXiv:1811.01848}} (\bibinfo{year}{2018}).
\newblock


\bibitem[Matsushima et~al\mbox{.}(2020)]%
        {matsushima2020deploymentefficient}
\bibfield{author}{\bibinfo{person}{Tatsuya Matsushima}, \bibinfo{person}{Hiroki Furuta}, \bibinfo{person}{Yutaka Matsuo}, \bibinfo{person}{Ofir Nachum}, {and} \bibinfo{person}{Shixiang Gu}.} \bibinfo{year}{2020}\natexlab{}.
\newblock \bibinfo{title}{Deployment-Efficient Reinforcement Learning via Model-Based Offline Optimization}.
\newblock
\newblock
\showeprint[arxiv]{2006.03647}~[cs.LG]


\bibitem[Maurera et~al\mbox{.}(2023)]%
        {maurera2023impression}
\bibfield{author}{\bibinfo{person}{Fernando B~P{\'e}rez Maurera}, \bibinfo{person}{Maurizio~Ferrari Dacrema}, \bibinfo{person}{Pablo Castells}, {and} \bibinfo{person}{Paolo Cremonesi}.} \bibinfo{year}{2023}\natexlab{}.
\newblock \showarticletitle{Impression-Aware Recommender Systems}.
\newblock \bibinfo{journal}{\emph{arXiv preprint arXiv:2308.07857}} (\bibinfo{year}{2023}).
\newblock


\bibitem[Rafailov et~al\mbox{.}(2021)]%
        {rafailov2020offline}
\bibfield{author}{\bibinfo{person}{Rafael Rafailov}, \bibinfo{person}{Tianhe Yu}, \bibinfo{person}{Aravind Rajeswaran}, {and} \bibinfo{person}{Chelsea Finn}.} \bibinfo{year}{2021}\natexlab{}.
\newblock \bibinfo{title}{Offline Reinforcement Learning from Images with Latent Space Models}.
\newblock , \bibinfo{numpages}{1154--1168}~pages.
\newblock


\bibitem[Rendle et~al\mbox{.}(2010)]%
        {rendle2010factorizing}
\bibfield{author}{\bibinfo{person}{Steffen Rendle}, \bibinfo{person}{Christoph Freudenthaler}, {and} \bibinfo{person}{Lars Schmidt-Thieme}.} \bibinfo{year}{2010}\natexlab{}.
\newblock \showarticletitle{Factorizing personalized markov chains for next-basket recommendation}. In \bibinfo{booktitle}{\emph{Proceedings of the 19th international conference on World wide web}}. \bibinfo{pages}{811--820}.
\newblock


\bibitem[Rigter et~al\mbox{.}(2022)]%
        {rigter2022rambo}
\bibfield{author}{\bibinfo{person}{Marc Rigter}, \bibinfo{person}{Bruno Lacerda}, {and} \bibinfo{person}{Nick Hawes}.} \bibinfo{year}{2022}\natexlab{}.
\newblock \showarticletitle{Rambo-rl: Robust adversarial model-based offline reinforcement learning}.
\newblock \bibinfo{journal}{\emph{Advances in neural information processing systems}}  \bibinfo{volume}{35} (\bibinfo{year}{2022}), \bibinfo{pages}{16082--16097}.
\newblock


\bibitem[Saito et~al\mbox{.}(2020)]%
        {DBLP:conf/wsdm/SaitoYNSN20}
\bibfield{author}{\bibinfo{person}{Yuta Saito}, \bibinfo{person}{Suguru Yaginuma}, \bibinfo{person}{Yuta Nishino}, \bibinfo{person}{Hayato Sakata}, {and} \bibinfo{person}{Kazuhide Nakata}.} \bibinfo{year}{2020}\natexlab{}.
\newblock \showarticletitle{Unbiased Recommender Learning from Missing-Not-At-Random Implicit Feedback}. In \bibinfo{booktitle}{\emph{{WSDM}}}. \bibinfo{publisher}{{ACM}}, \bibinfo{pages}{501--509}.
\newblock


\bibitem[Sarwar et~al\mbox{.}(2001)]%
        {sarwar2001itemcf}
\bibfield{author}{\bibinfo{person}{Badrul Sarwar}, \bibinfo{person}{George Karypis}, \bibinfo{person}{Joseph Konstan}, {and} \bibinfo{person}{John Riedl}.} \bibinfo{year}{2001}\natexlab{}.
\newblock \showarticletitle{Item-based collaborative filtering recommendation algorithms}. In \bibinfo{booktitle}{\emph{WWW}}. \bibinfo{pages}{285--295}.
\newblock


\bibitem[Schnabel et~al\mbox{.}(2016)]%
        {ips_rec}
\bibfield{author}{\bibinfo{person}{Tobias Schnabel}, \bibinfo{person}{Adith Swaminathan}, \bibinfo{person}{Ashudeep Singh}, \bibinfo{person}{Navin Chandak}, {and} \bibinfo{person}{Thorsten Joachims}.} \bibinfo{year}{2016}\natexlab{}.
\newblock \showarticletitle{Recommendations as treatments: Debiasing learning and evaluation}. In \bibinfo{booktitle}{\emph{international conference on machine learning}}. PMLR, \bibinfo{pages}{1670--1679}.
\newblock


\bibitem[Schrittwieser et~al\mbox{.}(2021)]%
        {schrittwieser2021online}
\bibfield{author}{\bibinfo{person}{Julian Schrittwieser}, \bibinfo{person}{Thomas Hubert}, \bibinfo{person}{Amol Mandhane}, \bibinfo{person}{Mohammadamin Barekatain}, \bibinfo{person}{Ioannis Antonoglou}, {and} \bibinfo{person}{David Silver}.} \bibinfo{year}{2021}\natexlab{}.
\newblock \showarticletitle{Online and offline reinforcement learning by planning with a learned model}.
\newblock \bibinfo{journal}{\emph{Advances in Neural Information Processing Systems}}  \bibinfo{volume}{34} (\bibinfo{year}{2021}), \bibinfo{pages}{27580--27591}.
\newblock


\bibitem[Siegel et~al\mbox{.}(2020)]%
        {siegel2020keep}
\bibfield{author}{\bibinfo{person}{Noah~Y Siegel}, \bibinfo{person}{Jost~Tobias Springenberg}, \bibinfo{person}{Felix Berkenkamp}, \bibinfo{person}{Abbas Abdolmaleki}, \bibinfo{person}{Michael Neunert}, \bibinfo{person}{Thomas Lampe}, \bibinfo{person}{Roland Hafner}, \bibinfo{person}{Nicolas Heess}, {and} \bibinfo{person}{Martin Riedmiller}.} \bibinfo{year}{2020}\natexlab{}.
\newblock \showarticletitle{Keep doing what worked: Behavioral modelling priors for offline reinforcement learning}.
\newblock \bibinfo{journal}{\emph{arXiv preprint arXiv:2002.08396}} (\bibinfo{year}{2020}).
\newblock


\bibitem[Sun et~al\mbox{.}(2019)]%
        {sun2019bert4rec}
\bibfield{author}{\bibinfo{person}{Fei Sun}, \bibinfo{person}{Jun Liu}, \bibinfo{person}{Jian Wu}, \bibinfo{person}{Changhua Pei}, \bibinfo{person}{Xiao Lin}, \bibinfo{person}{Wenwu Ou}, {and} \bibinfo{person}{Peng Jiang}.} \bibinfo{year}{2019}\natexlab{}.
\newblock \showarticletitle{BERT4Rec: Sequential recommendation with bidirectional encoder representations from transformer}. In \bibinfo{booktitle}{\emph{Proceedings of the 28th ACM international conference on information and knowledge management}}. \bibinfo{pages}{1441--1450}.
\newblock


\bibitem[Swazinna et~al\mbox{.}(2021)]%
        {swazinna2021overcoming}
\bibfield{author}{\bibinfo{person}{Phillip Swazinna}, \bibinfo{person}{Steffen Udluft}, {and} \bibinfo{person}{Thomas Runkler}.} \bibinfo{year}{2021}\natexlab{}.
\newblock \bibinfo{title}{Overcoming Model Bias for Robust Offline Deep Reinforcement Learning}.
\newblock
\newblock
\showeprint[arxiv]{2008.05533}~[cs.LG]


\bibitem[Vaswani(2017)]%
        {vaswani2017attention}
\bibfield{author}{\bibinfo{person}{A Vaswani}.} \bibinfo{year}{2017}\natexlab{}.
\newblock \showarticletitle{Attention is all you need}.
\newblock \bibinfo{journal}{\emph{Advances in Neural Information Processing Systems}} (\bibinfo{year}{2017}).
\newblock


\bibitem[Wang et~al\mbox{.}(2020)]%
        {wang2020reinforced}
\bibfield{author}{\bibinfo{person}{Xiang Wang}, \bibinfo{person}{Yaokun Xu}, \bibinfo{person}{Xiangnan He}, \bibinfo{person}{Yixin Cao}, \bibinfo{person}{Meng Wang}, {and} \bibinfo{person}{Tat-Seng Chua}.} \bibinfo{year}{2020}\natexlab{}.
\newblock \showarticletitle{Reinforced negative sampling over knowledge graph for recommendation}. In \bibinfo{booktitle}{\emph{Proceedings of The Web Conference 2020}}. \bibinfo{pages}{99--109}.
\newblock


\bibitem[Wang et~al\mbox{.}(2019)]%
        {wang2019doubly}
\bibfield{author}{\bibinfo{person}{Xiaojie Wang}, \bibinfo{person}{Rui Zhang}, \bibinfo{person}{Yu Sun}, {and} \bibinfo{person}{Jianzhong Qi}.} \bibinfo{year}{2019}\natexlab{}.
\newblock \showarticletitle{Doubly robust joint learning for recommendation on data missing not at random}. In \bibinfo{booktitle}{\emph{International Conference on Machine Learning}}. PMLR, \bibinfo{pages}{6638--6647}.
\newblock


\bibitem[Wang et~al\mbox{.}(2022a)]%
        {wang2022surrogate}
\bibfield{author}{\bibinfo{person}{Yuyan Wang}, \bibinfo{person}{Mohit Sharma}, \bibinfo{person}{Can Xu}, \bibinfo{person}{Sriraj Badam}, \bibinfo{person}{Qian Sun}, \bibinfo{person}{Lee Richardson}, \bibinfo{person}{Lisa Chung}, \bibinfo{person}{Ed~H Chi}, {and} \bibinfo{person}{Minmin Chen}.} \bibinfo{year}{2022}\natexlab{a}.
\newblock \showarticletitle{Surrogate for long-term user experience in recommender systems}. In \bibinfo{booktitle}{\emph{Proceedings of the 28th ACM SIGKDD conference on knowledge discovery and data mining}}. \bibinfo{pages}{4100--4109}.
\newblock


\bibitem[Wang et~al\mbox{.}(2022b)]%
        {wang2022lantentconfounders}
\bibfield{author}{\bibinfo{person}{Zhenlei Wang}, \bibinfo{person}{Shiqi Shen}, \bibinfo{person}{Zhipeng Wang}, \bibinfo{person}{Bo Chen}, \bibinfo{person}{Xu Chen}, {and} \bibinfo{person}{Ji-Rong Wen}.} \bibinfo{year}{2022}\natexlab{b}.
\newblock \showarticletitle{Unbiased sequential recommendation with latent confounders}. In \bibinfo{booktitle}{\emph{WWW}}. \bibinfo{pages}{2195--2204}.
\newblock


\bibitem[Xin et~al\mbox{.}(2020)]%
        {xin2020self}
\bibfield{author}{\bibinfo{person}{Xin Xin}, \bibinfo{person}{Alexandros Karatzoglou}, \bibinfo{person}{Ioannis Arapakis}, {and} \bibinfo{person}{Joemon~M Jose}.} \bibinfo{year}{2020}\natexlab{}.
\newblock \showarticletitle{Self-supervised reinforcement learning for recommender systems}. In \bibinfo{booktitle}{\emph{Proceedings of the 43rd International ACM SIGIR conference on research and development in Information Retrieval}}. \bibinfo{pages}{931--940}.
\newblock


\bibitem[Xu et~al\mbox{.}(2022)]%
        {DBLP:conf/cikm/Xu0CDW22}
\bibfield{author}{\bibinfo{person}{Chen Xu}, \bibinfo{person}{Jun Xu}, \bibinfo{person}{Xu Chen}, \bibinfo{person}{Zhenghua Dong}, {and} \bibinfo{person}{Ji-Rong Wen}.} \bibinfo{year}{2022}\natexlab{}.
\newblock \showarticletitle{Dually Enhanced Propensity Score Estimation in Sequential Recommendation}. In \bibinfo{booktitle}{\emph{CIKM}}. \bibinfo{pages}{2260--2269}.
\newblock


\bibitem[Yang et~al\mbox{.}(2024)]%
        {yang2024debiasing}
\bibfield{author}{\bibinfo{person}{Jiyuan Yang}, \bibinfo{person}{Yue Ding}, \bibinfo{person}{Yidan Wang}, \bibinfo{person}{Pengjie Ren}, \bibinfo{person}{Zhumin Chen}, \bibinfo{person}{Fei Cai}, \bibinfo{person}{Jun Ma}, \bibinfo{person}{Rui Zhang}, \bibinfo{person}{Zhaochun Ren}, {and} \bibinfo{person}{Xin Xin}.} \bibinfo{year}{2024}\natexlab{}.
\newblock \showarticletitle{Debiasing Sequential Recommenders through Distributionally Robust Optimization over System Exposure}. In \bibinfo{booktitle}{\emph{Proceedings of the 17th ACM International Conference on Web Search and Data Mining}}. \bibinfo{pages}{882--890}.
\newblock


\bibitem[Yang et~al\mbox{.}(2018)]%
        {Yang2018unbiasevaluatuion}
\bibfield{author}{\bibinfo{person}{Longqi Yang}, \bibinfo{person}{Yin Cui}, \bibinfo{person}{Yuan Xuan}, \bibinfo{person}{Chenyang Wang}, \bibinfo{person}{Serge Belongie}, {and} \bibinfo{person}{Deborah Estrin}.} \bibinfo{year}{2018}\natexlab{}.
\newblock \showarticletitle{Unbiased offline recommender evaluation for missing-not-at-random implicit feedback}. In \bibinfo{booktitle}{\emph{Proceedings of the 12th ACM conference on recommender systems}}. \bibinfo{pages}{279--287}.
\newblock


\bibitem[Yu et~al\mbox{.}(2021)]%
        {yu2021combo}
\bibfield{author}{\bibinfo{person}{Tianhe Yu}, \bibinfo{person}{Aviral Kumar}, \bibinfo{person}{Rafael Rafailov}, \bibinfo{person}{Aravind Rajeswaran}, \bibinfo{person}{Sergey Levine}, {and} \bibinfo{person}{Chelsea Finn}.} \bibinfo{year}{2021}\natexlab{}.
\newblock \showarticletitle{Combo: Conservative offline model-based policy optimization}.
\newblock \bibinfo{journal}{\emph{Advances in neural information processing systems}}  \bibinfo{volume}{34} (\bibinfo{year}{2021}), \bibinfo{pages}{28954--28967}.
\newblock


\bibitem[Yu et~al\mbox{.}(2020)]%
        {yu2020mopo}
\bibfield{author}{\bibinfo{person}{Tianhe Yu}, \bibinfo{person}{Garrett Thomas}, \bibinfo{person}{Lantao Yu}, \bibinfo{person}{Stefano Ermon}, \bibinfo{person}{James~Y Zou}, \bibinfo{person}{Sergey Levine}, \bibinfo{person}{Chelsea Finn}, {and} \bibinfo{person}{Tengyu Ma}.} \bibinfo{year}{2020}\natexlab{}.
\newblock \showarticletitle{Mopo: Model-based offline policy optimization}.
\newblock \bibinfo{journal}{\emph{Advances in Neural Information Processing Systems}}  \bibinfo{volume}{33} (\bibinfo{year}{2020}), \bibinfo{pages}{14129--14142}.
\newblock


\bibitem[Yuan et~al\mbox{.}(2019)]%
        {nextitnet}
\bibfield{author}{\bibinfo{person}{Fajie Yuan}, \bibinfo{person}{Alexandros Karatzoglou}, \bibinfo{person}{Ioannis Arapakis}, \bibinfo{person}{Joemon~M Jose}, {and} \bibinfo{person}{Xiangnan He}.} \bibinfo{year}{2019}\natexlab{}.
\newblock \showarticletitle{A Simple Convolutional Generative Network for Next Item Recommendation}. In \bibinfo{booktitle}{\emph{WSDM}}. ACM, \bibinfo{pages}{582--590}.
\newblock


\bibitem[Yuan et~al\mbox{.}(2022)]%
        {yuan2022tenrec}
\bibfield{author}{\bibinfo{person}{Guanghu Yuan}, \bibinfo{person}{Fajie Yuan}, \bibinfo{person}{Yudong Li}, \bibinfo{person}{Beibei Kong}, \bibinfo{person}{Shujie Li}, \bibinfo{person}{Lei Chen}, \bibinfo{person}{Min Yang}, \bibinfo{person}{Chenyun Yu}, \bibinfo{person}{Bo Hu}, \bibinfo{person}{Zang Li}, {et~al\mbox{.}}} \bibinfo{year}{2022}\natexlab{}.
\newblock \showarticletitle{Tenrec: A large-scale multipurpose benchmark dataset for recommender systems}.
\newblock \bibinfo{journal}{\emph{Advances in Neural Information Processing Systems}}  \bibinfo{volume}{35} (\bibinfo{year}{2022}), \bibinfo{pages}{11480--11493}.
\newblock


\bibitem[Zhan et~al\mbox{.}(2022)]%
        {zhan2022modelbased}
\bibfield{author}{\bibinfo{person}{Xianyuan Zhan}, \bibinfo{person}{Xiangyu Zhu}, {and} \bibinfo{person}{Haoran Xu}.} \bibinfo{year}{2022}\natexlab{}.
\newblock \bibinfo{title}{Model-Based Offline Planning with Trajectory Pruning}.
\newblock
\newblock
\showeprint[arxiv]{2105.07351}~[cs.AI]


\bibitem[Zhao et~al\mbox{.}(2018)]%
        {zhao2018deep}
\bibfield{author}{\bibinfo{person}{Xiangyu Zhao}, \bibinfo{person}{Long Xia}, \bibinfo{person}{Liang Zhang}, \bibinfo{person}{Zhuoye Ding}, \bibinfo{person}{Dawei Yin}, {and} \bibinfo{person}{Jiliang Tang}.} \bibinfo{year}{2018}\natexlab{}.
\newblock \showarticletitle{Deep reinforcement learning for page-wise recommendations}. In \bibinfo{booktitle}{\emph{Proceedings of the 12th ACM conference on recommender systems}}. \bibinfo{pages}{95--103}.
\newblock


\bibitem[Zhou et~al\mbox{.}(2020)]%
        {zhou2020s3rec}
\bibfield{author}{\bibinfo{person}{Kun Zhou}, \bibinfo{person}{Hui Wang}, \bibinfo{person}{Wayne~Xin Zhao}, \bibinfo{person}{Yutao Zhu}, \bibinfo{person}{Sirui Wang}, \bibinfo{person}{Fuzheng Zhang}, \bibinfo{person}{Zhongyuan Wang}, {and} \bibinfo{person}{Ji-Rong Wen}.} \bibinfo{year}{2020}\natexlab{}.
\newblock \showarticletitle{S3-rec: Self-supervised learning for sequential recommendation with mutual information maximization}. In \bibinfo{booktitle}{\emph{Proceedings of the 29th ACM international conference on information \& knowledge management}}. \bibinfo{pages}{1893--1902}.
\newblock


\end{thebibliography}

\appendix

\end{document}